\documentclass[transmag]{IEEEtran}
\usepackage{latexsym}
\usepackage{graphicx}
\usepackage{amsfonts,amssymb,amsmath}
\usepackage{hyperref}

\usepackage[T1]{fontenc}
\usepackage{cite}
\usepackage{subcaption}

\usepackage{amsthm}
\usepackage{overpic}
\usepackage{steinmetz}
\usepackage{array}
\usepackage{url}
\usepackage{color}

\theoremstyle{plain}

\newtheorem{lemma}{Lemma}
\newtheorem{example}{Example}
\newtheorem{corollary}{Corollary}
\newtheorem{remark}{Remark}
\newtheorem{proposition}{Proposition}
\newtheorem{assumption}{Assumption}

\newcommand{\vect}[1]{\mathbf{#1}}

\def\diag{\mathrm{diag}}

\def\Htran{\mbox{\tiny $\mathrm{H}$}}
\def\Ttran{\mbox{\tiny $\mathrm{T}$}}
\def\CN{\mathcal{N}_{\mathbb{C}}} 

\def\Ptx{P_{\mathrm{tx}}}
\def\Prelay{P_{\mathrm{relay}}}
\def\Prx{P_{\mathrm{rx}}}
\def\mod{\mathrm{mod}}

\def\BibTeX{{\rm B\kern-.05em{\sc i\kern-.025em b}\kern-.08em T\kern-.1667em\lower.7ex\hbox{E}\kern-.125emX}}
\begin{document}

\title{Power Scaling Laws and Near-Field Behaviors of Massive MIMO and Intelligent Reflecting Surfaces}

\author{Emil Bj{\"o}rnson, \IEEEmembership{Senior Member, IEEE}, Luca Sanguinetti,
\IEEEmembership{Senior Member, IEEE}
\thanks{A preliminary version of this paper was presented at the IEEE CAMSAP 2019 \cite{Bjornson2019f}.}
\thanks{E. Bj{\"o}rnson was supported by ELLIIT and the Wallenberg AI, Autonomous Systems and Software Program (WASP). L. Sanguinetti was partially supported by the University of Pisa under the PRA 2018-2019 Research Project CONCEPT, and by the Italian Ministry of Education and Research (MIUR) in the framework of the CrossLab project (Departments of Excellence).}
\thanks{E.~Bj\"ornson is with the Department of Electrical Engineering (ISY), Link\"{o}ping University, 58183 Link\"{o}ping, Sweden (emil.bjornson@liu.se).}
\thanks{L.~Sanguinetti is with the Dipartimento di Ingegneria dell'Informazione, University of Pisa, 56122 Pisa, Italy (luca.sanguinetti@unipi.it).}}

\IEEEtitleabstractindextext{\begin{abstract}The use of large arrays might be the solution to the capacity problems in wireless communications. The signal-to-noise ratio (SNR) grows linearly with the number of array elements $N$ when using Massive MIMO receivers and half-duplex relays. Moreover, intelligent reflecting surfaces (IRSs) have recently attracted attention since these can relay signals to achieve an SNR that grows as $N^2$, which seems like a major benefit. In this paper, we use a deterministic propagation model for a planar array of arbitrary size, to demonstrate that the mentioned SNR behaviors, and associated power scaling laws, only apply in the far-field. They cannot be used to study the regime where $N\to\infty$. We derive an exact channel gain expression that captures three essential near-field behaviors and use it to revisit the power scaling laws. We derive new finite asymptotic SNR limits but also conclude that these are unlikely to be approached in practice. We further prove that an IRS-aided setup cannot achieve a higher SNR than an equal-sized Massive MIMO setup, despite its faster SNR growth. We quantify analytically how much larger the IRS must be to achieve the same SNR. Finally, we show that an optimized IRS does not behave as an ``anomalous'' mirror but can vastly outperform that benchmark.
\end{abstract}

}

\maketitle

\section{Introduction}

Massive multiple-input multiple-output (mMIMO) is the key physical layer technology in 5G \cite{Parkvall2017a}. In a nutshell, mMIMO uses a base station with many antennas (e.g., $\geq 64$) to deliver large array gains and perform spatial multiplexing of many users on the same time-frequency resource \cite{Marzetta2010a,massivemimobook,Sanguinetti2019az}. In this way, the spectral efficiency (SE) can be increased by, at least, an order of magnitude compared to 4G and mmWave communications can be enabled in mobile networks. 
Due to the success of mMIMO, it is expected that beyond 5G systems will make use of even larger arrays and wider spectrum ranges \cite{Bjornson2019d,Rajatheva2020a}. The arrays can either consist of active or passive elements, and both cases are considered in this paper.

The active arrays are essentially mMIMO transceivers but with many more antenna elements than what is considered in 5G (i.e., much more than 64). To make this clear, the research community has recently used new names to describe this category; for example, \emph{large intelligent surfaces} \cite{Hu2018a}, \emph{extremely large aperture arrays}  \cite{Amiri2018a}, and \emph{holographic MIMO} \cite{Pizzo2019a}. However, in this paper, we will simply refer to it as mMIMO since asymptotically large arrays have been analyzed since the inception of mMIMO \cite{Marzetta2010a,Ngo2013a,Hoydis2013a}.

The passive arrays are large metasurfaces \cite{Liang2015a,PhysRevX.8.011036} that are deployed somewhere in the propagation environment (in between the transmitter and receiver) to support the transmission from a source to a destination by creating and shaping additional propagation paths. A metasurface consists of many sub-wavelength-sized elements that each acts as a diffuse scatterer \cite{Liang2015a} but with the special feature of being able to adjust the phase (i.e., time delay) and polarization. By controlling the phase-shifts of the individual elements, the metasurface can ``reflect'' an incident wave as a beam in the desired direction \cite{PhysRevX.8.011036}; the physics is the same as for beamforming using a phased array, except that the array then generates the signal locally. 
The concept of real-time controllable metasurfaces has recently received much attention in the communications society and is called \emph{intelligent reflecting surface (IRS)} \cite{Wu2018a,Wu2020a}, \emph{software-controlled metasurface} \cite{Liaskos2018a,Renzo2019a}, and \emph{reconfigurable intelligent surface} \cite{Huang2018a,Basar2019a,Renzo2020b}. We will call it IRS in this paper and consider the case when the metasurface is used as a relay that reflects the signal from the source towards the destination in an effort to maximize the signal-to-noise ratio (SNR). Using conventional relaying terminology \cite{Dohler2010a,Bjornson2020a}, an IRS is a transparent relay since it processes the received signal in the analog domain and operates in full duplex since the signals are received and reflected simultaneously. Different from conventional relays, the signal is not amplified in the IRS but it instead improves the SNR by capitalizing on the array power gain achieved when having a large surface.
We will compare the IRS with the use of a conventional half-duplex mMIMO relay, which is also deployed in between a source and destination to improve the propagation conditions; see \cite{Ngo2014c,Yang2015a,Kong2018a} and reference therein for prior work on mMIMO relays.
We refer to \cite{Liaskos2018a,Renzo2020b,Wu2020a,Rajatheva2020a} for an overview of other prospective use cases of the IRS technology beyond relaying, including how it can be used in conjunction with other technologies.

A fundamental benefit of using large arrays is that the SNR grows with the number of elements $N$. In mMIMO setups, the SNR is proportional to $N$ if optimal beamforming is applied \cite{Ngo2013a,Hoydis2013a,Ngo2014c,Yang2015a,Kong2018a}. This implies that the transmit power needed to achieve a target SNR value during data transmission reduces as $1/N$, which is a so-called asymptotic power scaling law.\footnote{If one also reduces the transmit power in the channel acquisition phase, the power scaling law changes; we refer to \cite{Ngo2013a,Hoydis2013a,massivemimobook} for details.} In contrast, the SNR grows as $N^2$ when using an IRS that is optimally configured \cite{Wu2018a}. Hence, a more aggressive power scaling law can be formulated where the transmit power is reduced as $1/N^2$ \cite{Wu2018a,Wu2019a,Basar2019a}.

A main limitation of the aforementioned SNR analyses and power scaling laws is that they are derived under an implicit assumption of far-field operation. When considering arrays, the far-field refers to the propagation range at which the direction and channel gain are approximately the same from all elements in the array to the transmitting/receiving antenna. This is different from the Fraunhofer distance, which measures the radiative far-field from a single antenna element and will not be considered in this paper.
Since the array size grows with $N$, we will inevitably operate in the array's near-field as $N \to \infty$. The near-field behavior of the SNR is uncertain since different papers have put forward different hypotheses.
For example, several papers have studied the IRS behavior in the far- and near-field assuming that it operates as a specular reflector (also called an ``anomalous'' mirror)\cite{Basar2019a,Yildirim2019a,Renzo2020a} and made parallels to geometrical physics to support this assumption.

The conference version \cite{Bjornson2019f} of this paper was the first attempt to mathematically derive the near-field behavior in both the mMIMO and IRS setups, but the results are approximate since they relied on the propagation model from \cite{Hu2018a} that neglects polarization effects.
The mismatch between the polarization of an antenna and of the incident wave is approximately the same for all antennas in the far-field, thus one can compute the SNR without polarization and then multiply with a coefficient accounting for the mismatch loss \cite[Sec.~7.4]{massivemimobook}.
The situation is more complicated in the near-field (i.e., for large arrays) where the incident wave is arriving from distinctly different angular directions to different elements, thus one must model the polarization on an element-by-element basis to obtain accurate results.
Recently, \cite{Tang2019a,Ellingson2019a,Garcia2019a} provided numerical studies and discussions regarding the near-field behavior, but the results are approximate since polarization is neglected in these works. In \cite{Tang2019a,Garcia2019a}, the effective areas of the elements are also assumed constant in the array, which is not the case in the near-field since the elements are observed from distinctly different angles. This is an additional source of approximation errors. Nevertheless, the experimental results presented in \cite{Tang2019a} show that there are finite-sized setups where the approximate formulas are matching quite well with measurements. However, we will show later that none of these prior analytical results can be used to characterize the asymptotic limit where $N \to \infty$.

\subsection{Contributions}

We derive a novel closed-form expression for the channel gain when communicating between a single-antenna device and a planar array of arbitrary size, by taking the varying distances to the elements, 
polarization mismatches, and effective areas into account. We demonstrate that it is necessary to utilize this model to rigorously study the signal propagation in the array's near-field and the asymptotic limits where the array dimensions grow large, because the approximate models in previous work give different results.
We use the derived expression to mathematically derive the near-field and far-field behaviors in three key setups: conventional mMIMO, half-duplex mMIMO relaying, and IRS-aided communications. In particular, we explain under which conditions the SNR grows with $N$ in the ways described above, and when we instead need to consider the alternative near-field behaviors that we establish. The analysis shows that the far-field approximation is accurate when the distance to the array is larger than its height/width, which holds in many practical scenarios but will not hold when studying the asymptotic limit when $N\to \infty$. We derive new power scaling laws that are asymptotically accurate.
Furthermore, we prove that an IRS cannot achieve a higher SNR than any of the mMIMO setups when the array sizes are equal, despite the fact that the SNR in the IRS setup grows as $N^2$ in the far-field. We derive closed-form expressions for how large an IRS must be to beat conventional mMIMO or half-duplex mMIMO relaying. We provide a geometric interpretation of an SNR-maximizing IRS, which is different from the specular reflector scenario that has been assumed in some prior work. While the main theory is developed for a free-space line-of-sight setup, we also extend the results to consider arbitrary deterministic channel models.

\subsection{Outline}

Preliminaries on signal propagation and array gains are provided in Section~\ref{sec:preliminaries}. The channel between a single-antenna transmitter and a planar antenna array of arbitrary size is derived in Section~\ref{planar_arrays}, assuming a free-space line-of-sight scenario. The asymptotic limits are derived and the asymptotic deficiencies of previously used models are exemplified.
We define the system models and achievable spectral efficiencies of three different setups in Section~\ref{sec:three-setups}: conventional mMIMO, half-duplex mMIMO relaying, and IRS-aided communications. The power scaling laws and near/far-field behaviors of these setups are uncovered in Section~\ref{sec:power-scaling-laws}, by utilizing the results from Section~\ref{planar_arrays}.
The case with different array sizes is studied in Section~\ref{sec:how-many-elements}, to quantify how much larger an IRS must be to match the spectral efficiencies achieved by the other setups.
Next, in Section~\ref{sec:geometric-interpretation}, we provide a new geometric interpretation of an IRS that is configured to maximize the SNR. Section~\ref{sec:extensions} discusses the extension to more general channel models.
Finally, the main results and conclusions are summarized in Section~\ref{sec:conclusion}.

\subsection{Reproducible Research}

The simulation results can be reproduced using code available at: \url{https://github.com/emilbjornson/near-field-behavior}

\subsection{Notation}

Boldface lowercase letters, $\vect{x}$, denote column vectors and boldface uppercase letters, $\vect{X}$, denote matrices. 
The superscripts $^{\Ttran}$, $^*$, and $^{\Htran}$ denote transpose, conjugate, and conjugate transpose, respectively. 
 The $n \times n$ identity matrix is $\vect{I}_n$, $\mod(\cdot,\cdot)$ indicates the modulo operation, and $\lfloor \cdot \rfloor$ rounds to the argument to the closest smaller integer. The multi-variate circularly symmetric complex Gaussian distribution with covariance matrix $\vect{R}$ is denoted $\CN(\vect{0},\vect{R})$. We define $||{\bf x}||$ the Frobenius norm of vector ${\bf x}$.

\section{Preliminaries}
\label{sec:preliminaries}

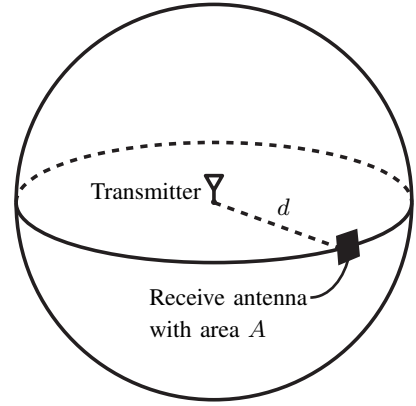
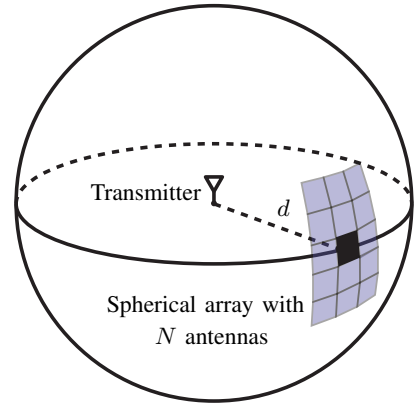
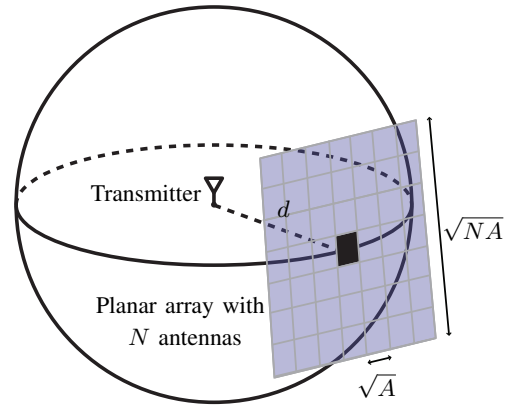
\begin{figure}[t!]
        \centering
        \begin{subfigure}[b]{\columnwidth} \centering 
	\begin{overpic}[width=.8\columnwidth,tics=10]{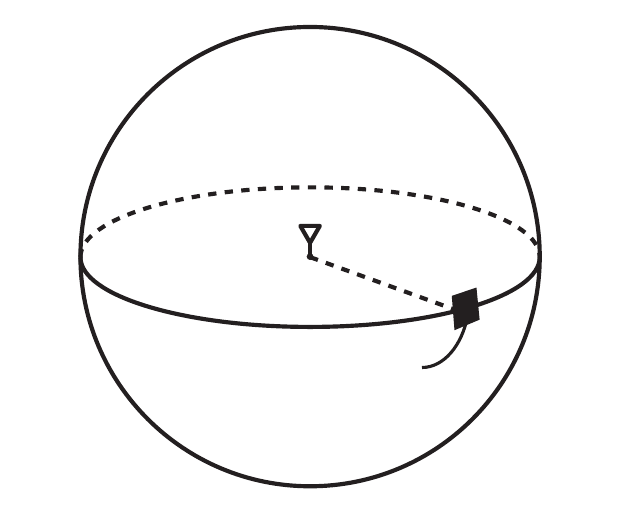}
  \put (27,43) {\small Transmitter}
 \put (62,40) {\small $d$}
  \put (38,23) {\small Receive antenna}
     \put (38,17) {\small with area $A$}
\end{overpic}  \vspace{-2mm}
                \caption{One receive antenna with area $A$.} 
                \label{fig:sphere}
        \end{subfigure}\\
        \begin{subfigure}[b]{\columnwidth} \centering  \vspace{+2mm}
	\begin{overpic}[width=.8\columnwidth,tics=10]{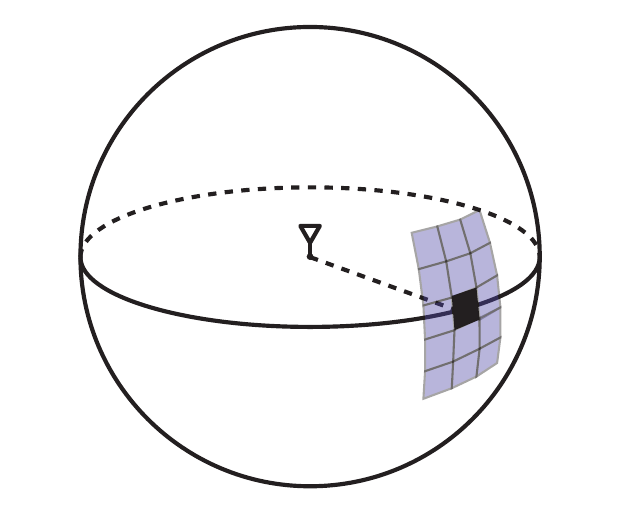}
  \put (27,43) {\small Transmitter}
 \put (62,40) {\small $d$}
  \put (30,22) {\small Spherical array with}
     \put (39,16) {\small $N$ antennas}
\end{overpic}  \vspace{-2mm}
                \caption{Spherical array with $N$ equal-sized receive antennas.} 
                \label{fig:sphere2}
        \end{subfigure} 
        \begin{subfigure}[b]{\columnwidth} \centering  \vspace{+2mm}
	\begin{overpic}[width=.8\columnwidth,tics=10]{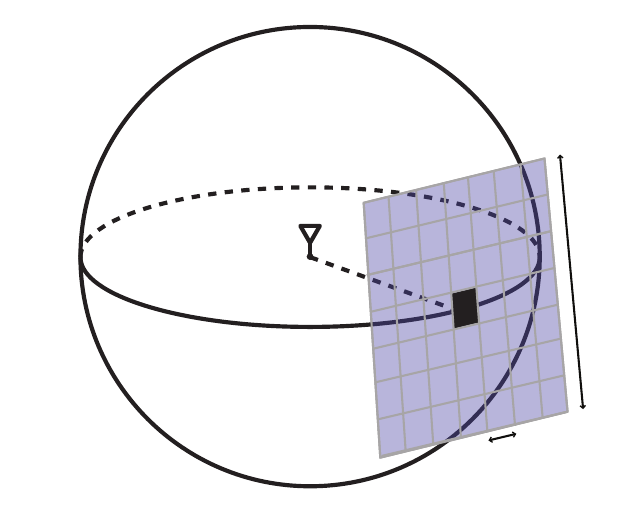}
  \put (27,43) {\small Transmitter}
 \put (62,40) {\small $d$}
  \put (28,22) {\small Planar array with}
     \put (34,16) {\small $N$ antennas}
   \put (93,36) {\small $\sqrt{NA}$}
   \put (77,6.5) {\small $\sqrt{A}$}
\end{overpic}  \vspace{-2mm}
                \caption{Planar array with $\sqrt{N} \times \sqrt{N}$ equal-sized receive antennas.} 
                \label{fig:sphere3}
        \end{subfigure} 
        \caption{Examples of basic antenna scenarios.}
        \label{fig:sphere12_examples}
\end{figure}

This paper analyzes the wireless propagation when using arrays of different sizes and transmitters/receivers at different distances. 
We begin by considering the free-space propagation scenario shown in Fig.~\ref{fig:sphere}, where an ideal isotropic transmit antenna sends a signal to a receive antenna located at distance $d$. Assume that the receive antenna is lossless, has an (effective) area $A$ perpendicular to the direction of propagation, and has a polarization perfectly matching that of the transmitted signal. Then, from Friis' formula \cite{friis1946note}, the received power is
\begin{equation} \label{eq:received_power_SISO}
\Prx =  \frac{A}{4\pi d^2} \Ptx
\end{equation}
where $\Ptx$ denotes the transmit power and the factor
\begin{equation} \label{eq:beta-definition}
\beta_d=\frac{A}{4\pi d^2}
\end{equation}
is the free-space channel gain, also known as pathloss. Note that this factor is given by the area $A$ of the receive antenna divided by the total surface area of a sphere with radius $d$. We use the subscript $d$ in $\beta_d$ to express that the channel gain is a function of $d$. Since the received power $\Prx$ can never be higher than the transmit power $\Ptx$ (due to the law of conservation of energy), it is evident that $\beta_d \in [0,1]$. In most cases, $\beta_d$ is much smaller than one, as we will now exemplify.

\begin{example} \label{example1}
If the receive antenna is isotropic, its area is $A= {\lambda^2}/{(4\pi)}$ where $\lambda = c/f$ is the wavelength and $c$ is the speed of light.
If the transmission has a carrier frequency of $f=3$\,\textrm{GHz}, then $\lambda = 0.1$\,\textrm{m}. For propagation distances $d\in [2.5, 25]$\,m, the channel gain $\beta_d$ ranges from $-50$\,dB to $-70$\,dB. If the carrier frequency is increased to $f=30$\,GHz, the antenna area becomes 100 times smaller and thus $\beta_d$ will instead range from  $-70$\,dB to $-90$\,dB for $d\in [2.5, 25]$\,m.
\end{example}

A way to increase the channel gain in \eqref{eq:beta-definition} is to make the receive antenna area  larger. In particular, we can deploy $N$ antennas of the same kind in an array. If they are deployed  on the sphere in  Fig.~\ref{fig:sphere} and are non-overlapping, and if each antenna has an orientation and polarization that match the locally received signal, the total received power is $N$ times the value in \eqref{eq:received_power_SISO}:
\begin{equation} \label{eq:received_power_SISO2}
\Prx^{\textrm{spheric-}N} = N \Prx = N \beta_d \Ptx.
\end{equation}
This setup is illustrated in Fig.~\ref{fig:sphere2} and the channel gain is $N \beta_d = \frac{NA}{4\pi d^2}$. Note that the channel gain is proportional to the total antenna area $NA$, thus one can achieve the same result with many physically small receive antennas or a few large antennas.
Clearly, no more than $NA = 4\pi d^2$ non-overlapping receive antennas can be deployed on the sphere in the way shown in Fig.~\ref{fig:sphere2}. In that case, 
$N \beta_d = 1$ and $\Prx^{\textrm{spheric-}N} = \Ptx$, so that all the transmitted power is received. Observe that very many antennas are needed to make this happen. Under the assumptions in Example~\ref{example1}, we need $10^4$ antennas to cover the entire sphere for $d= 2.5$\,m when communicating at $f=3$\,GHz, and $10^6$ antennas for $d= 25$\,m.
Both values increase by 100 times when communicating at $f=30$\,GHz since the area of an isotropic receive antenna becomes 100 times smaller.

The linear growth with $N$ in \eqref{eq:received_power_SISO2} is called an \emph{array gain} and is the key motivating factor behind mMIMO communications using antenna arrays with a large (possibly infinite) number of antennas. A common assumption in such systems is that the linear scaling holds true even in the asymptotic regime where $N\to \infty$\cite{Marzetta2010a,Jose2011b,Hoydis2013a,Ngo2013a,Yin2013a,Mueller2014b,BjornsonHS17}, which has been utilized to define properties such as channel hardening and favorable propagation, as well as studying the fundamental impact of pilot contamination.
This assumption is physically incorrect since the total channel gain would be higher than one, thereby invalidating the law of conservation of energy. On the other hand, the analysis above has shown that a very large number of antennas is needed to receive all the transmitted power. Hence, the linear scaling might hold in practical mMIMO communications, even if thousands of antennas are used and the propagation distance is short. The aim of the next section is to revisit the asymptotic regime with practical planar antenna arrays and prove under what conditions the linear scaling is inaccurate versus approximately correct. These results will be fundamental in Section~\ref{sec:power-scaling-laws} when studying the power scaling laws and near-field behaviors of different mMIMO systems and IRS-aided setups.

\section{Planar Antenna Arrays}\label{planar_arrays}

We now turn the attention to the planar array illustrated in Fig.~\ref{fig:sphere3}, particularly because such arrays are commonly used in practical mMIMO deployments \cite{Bjornson2019d}. The transmit antenna is at distance $d$ from the center of the array. For notational convenience, we make the following assumption that will be considered in the remainder of this paper.

\begin{assumption} \label{assumption1}
The planar array consists of $N$ antennas that each has area $A\leq (\lambda/4)^2$. The antennas have size $\sqrt{A} \times \sqrt{A}$ and are equally spaced on a $\sqrt{N} \times \sqrt{N}$ grid. The antennas are deployed edge-to-edge, thus the total area of the array is $NA$.  
\end{assumption}

These assumptions\footnote{Assumption~\ref{assumption1} restricts $N$ to be the square of an integer, but the analytical results of this paper only require a quadratic planar array with dimension $\sqrt{NA} \times \sqrt{NA}$. For a given array area $NA$, we can always adapt $A$ to make $N$ be the square of an integer.} are important when quantifying the channel gain, because the \emph{effective} area of each receive antenna will depend on its physical location and rotation, with respect to the direction of the transmitter. The physical area of each antenna is $A$ but the effective area seen from the transmitter varies.
If the receive antenna is fully perpendicular to the direction of propagation, then the effective area  also equals $A$. In any other case, the effective area is smaller than $A$.
The antenna gain in a particular direction is determined by the effective antenna area seen from transmitter and also by the polarization loss caused by having a rotated aperture \cite{Dardari2019a}.\footnote{Only the components of the field vectors that are perpendicular to the boresight of the antenna can be received, irrespective of whether linearly or circularly polarized signals are considered. When deriving the analytical results, we consider linear polarization along the Y direction but since we assume a square array, the results are rotationally invariant. Therefore, another choice of polarization will lead to the same end results, even if the individual elements will contribute differently.}
When the transmitter is in the near-field of the array, three fundamental properties must be taken into account:

\begin{enumerate}
\item The distances to the elements vary over the array; 

\item The effective antenna areas vary since the element are seen from different angles; 

\item The losses from polarization mismatch vary since the signals are received from different angles.
\end{enumerate}

\subsection{Exact Expression for the Channel Gain}

The following lemma extends the prior work in \cite{Dardari2019a} to the case when the transmitter and the receiver are arbitrarily located, thereby providing a general way of computing the channel gains to each of the $N$ elements of a planar array.\footnote{We disregard the mutual coupling effect in this paper, to focus on the asymptotic behaviors with ideal hardware.}

\begin{lemma} \label{lemma1}
Consider a lossless isotropic antenna located at $\vect{p}_t=(x_t,y_t,d)$ that transmits a signal that has polarization in the $Y$ direction when traveling in the $Z$ direction.
 The receive antenna is located in the $XY$-plane, is centered at $\vect{p}_n=(x_n,y_n,0)$, and has area $a \times a$. The free-space channel gain is upper bounded by
\begin{align} \notag
\zeta_{\vect{p}_t,\vect{p}_n,a} = \frac{1}{4\pi} \sum_{x \in \mathcal{X}_{t,n}  } \sum_{ y \in \mathcal{Y}_{t,n}  } &\left( \frac{\frac{xy}{d^2}}{3 \left(\frac{y^2}{d^2}+1\right)\sqrt{ \frac{x^2}{d^2}+\frac{y^2}{d^2}+1}} \right. \\& \left.
+ \frac{2}{3} \tan^{-1} \left(  \frac{\frac{xy}{d^2}}{\sqrt{ \frac{x^2}{d^2}+\frac{y^2}{d^2}+1}}
\right) \right) \label{eq:channel-gain-general-case}
\end{align}
where $\mathcal{X}_{t,n} = \{ a/2+x_n-x_t,a/2-x_n+x_t \}$ and $\mathcal{Y}_{t,n}  = \{ a/2+y_n-y_t,a/2-y_n+y_t \}$.

The upper bound is tight when the antenna area is sufficiently small compared to the wavelength: $a \le \lambda/4$.
\end{lemma}
\begin{IEEEproof}
The proof is given in Appendix~\ref{app:proof-lemma1} in two steps. The impact of the three fundamental properties mentioned above is clearly pointed out in \eqref{eq:pathloss-integral}.
\end{IEEEproof}

This lemma provides an upper bound on the channel gain by assuming the received signal's phase-variations are negligible over the antenna area, which is commonly assumed in the literature but is only a tight bound for sub-wavelength-sized antennas, as will be assumed in this paper.
We will use the general formula from Lemma~\ref{lemma1} later in the paper, particularly when analyzing an IRS-aided setup. However, we first notice that a compact expression for the channel gain and, thus, the total received power can be obtained when the transmitter is centered in front of the planar array.

\begin{corollary} \label{cor:alpha-expression}
Under Assumption~\ref{assumption1}, when the transmitter is centered in front of the planar array, the received power is \vspace{-2mm}
\begin{equation} \label{eq:received_power_planar}
\Prx^{\mathrm{planar}\textrm{-}N}  = \alpha_{d,N} \Ptx
\end{equation}
where the total channel gain is
\begin{align} \label{eq:alpha-expression} \nonumber
\alpha_{d,N} 
&=   \frac{N \beta_d}{3(N \beta_d \pi+1) \sqrt{2 N \beta_d \pi + 1}} \\ &+  \frac{2}{3\pi}\tan^{-1} \!\left( \frac{N \beta_d \pi}{ \sqrt{2N \beta_d \pi + 1}} \right) 
\end{align}
with $\beta_d$ given in \eqref{eq:beta-definition}.
\end{corollary}
\begin{IEEEproof}
This formula follows from Lemma~\ref{lemma1} by setting $x_t=y_t=0$, $x_n=y_n=0$, and $a=\sqrt{NA}$, in which case $\mathcal{X}_{t,n}  = \mathcal{Y}_{t,n}  = \{\sqrt{NA}/2, \sqrt{NA}/2\}$. Even if we might have $a > \lambda/4$, the expression of $\alpha_{d,N}$ is a tight upper bound on the true channel gain since the individual antennas satisfy $\sqrt{A} \leq \lambda/4$ and we are summing up their received powers. By replacing $d$ with $\sqrt{A/(4\pi \beta_d) }$ and rearranging the terms, we obtain \eqref{eq:alpha-expression} from \eqref{eq:channel-gain-general-case}.
\end{IEEEproof}

The channel gain in \eqref{eq:alpha-expression} 
is valid for arbitrarily large planar arrays, which is different from the models considered in \cite{Tang2019a,Garcia2019a,Ellingson2019a} that assume equal effective areas of all elements. The new expression supports the case when the transmitter is in the near-field of the array.\footnote{Note that we assume throughout this paper that $d \gg \lambda$, so the system does not operate in the reactive near-field of the transmit antenna (even if it is in the near-field of the array). In fact, this assumption was made in the proof of the expression in Lemma~\ref{lemma1}.} We will explore the far-field approximation and large-array limit appearing in the near-field.

\begin{remark} \label{remark:NA-scaling}
The exact expression in \eqref{eq:alpha-expression} depends on $N \beta_d$, thus it is the total array area $NA$ that matters and not the individual values of $N$ and $A$. Hence, the results in this paper hold for any frequency band and choice of individual antenna areas, as long as the total area is the same and $A\leq (\lambda/4)^2$ so Lemma~\ref{lemma1} provides a tight bound for the individual antennas. As the wavelength reduces, the area $A$  shrinks and then more elements are needed to fill the same total array area. In all the simulation figures, we are considering $\lambda = 0.1$\,m (3 GHz) when reporting the number of elements $N$, but the same behaviors appear in any frequency band under the condition that the total area $NA$ of the array is the same.
\end{remark}

\subsection{Far-field Approximation and Large-array Limit}

Suppose the planar array considered in Corollary~\ref{cor:alpha-expression} is in the far-field of the transmitter in the sense that $d \gg \sqrt{N A}$. In this case, $N\beta_d \pi +1 \approx 1$ and $ \sqrt{2N \beta_d \pi + 1} \approx 1$.
By using the first-order Taylor approximation $\tan^{-1}(x) \approx x$, which is tight when the argument is close to zero (as is the case when $N \beta_d \pi$ is small), it follows from \eqref{eq:received_power_planar} that
\begin{equation} \label{eq:received_power_planar-approx}
\Prx^{\mathrm{planar}\textrm{-}N}  \approx \left( \frac{N \beta_d}{3} + \frac{2}{3\pi} N \beta_d \pi \right) \Ptx = N \beta_d \Ptx
\end{equation}
which is equal to $\Prx^{\textrm{spheric-}N}$ in  \eqref{eq:received_power_SISO2}. Hence, for relatively small planar arrays, the received power is proportional to $N$. Both terms in \eqref{eq:alpha-expression} contribute to the result, but not equally much.

If $N$ grows large, the far-field approximation is no longer valid and we instead notice that as $N \to \infty$ it holds that
\begin{align}
 \frac{N \beta_d}{3(N \beta_d \pi+1) \sqrt{2 N \beta_d \pi + 1}}  &\to 0, \\
\tan^{-1} \!\left( \frac{N \beta_d \pi}{ \sqrt{2N \beta_d \pi + 1}} \right) &\to \frac{\pi}{2}.
\end{align}
 Hence, the received power in \eqref{eq:received_power_planar} saturates and has the asymptotic limit 
 \begin{equation} \label{eq:limit-planar-array}
 \Prx^{\mathrm{planar}\textrm{-}N} \to \frac{2}{3\pi} \frac{\pi}{2}  \Ptx
 = \frac{\Ptx}{3} \quad \textrm{as} \,\,N\to \infty. 
  \end{equation}
  This value satisfies the law of conservation of energy since only one third of the transmitted power is received. An intuitive explanation for why the limit is finite, although the array is infinitely large, is that each new receive antenna is deployed further away from the transmitter; the effective area (perpendicularly to the direction of propagation) becomes gradually smaller and the polarization loss also increases. 
  
\begin{figure}[t!]
	\centering 
	\begin{overpic}[width=1.07\columnwidth,tics=10]{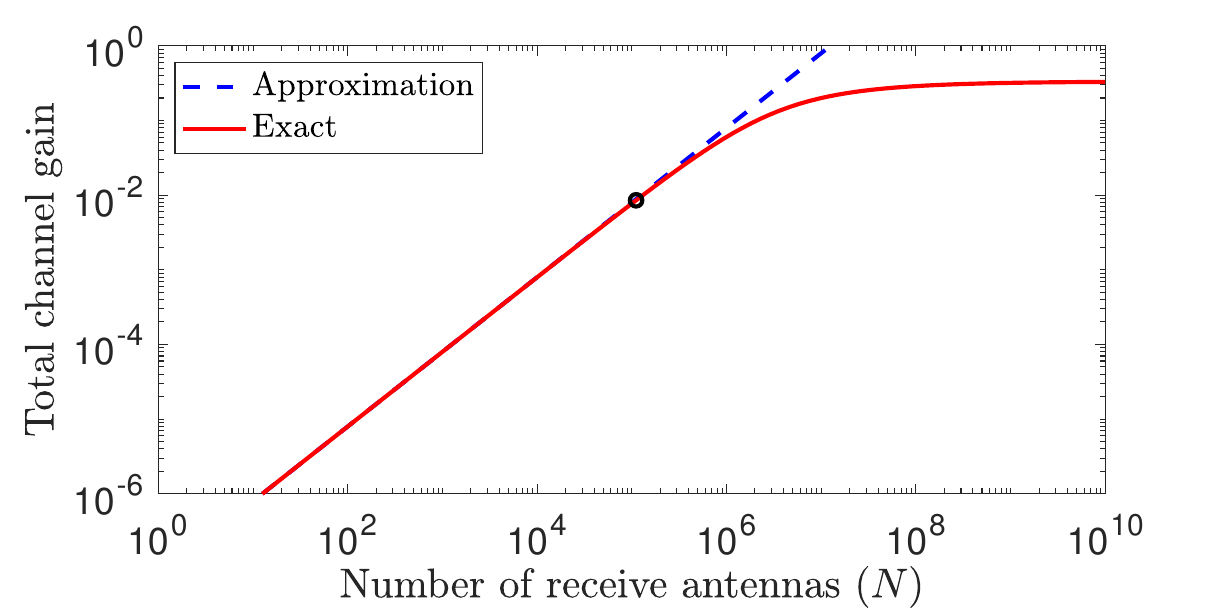}
	\put(44.5,25){\small Point given by the rule-of-thumb}
	\put(44.5,21){\small (where the far-field ends)}
	\put(58,28){\vector(-1, 1){5}}
\end{overpic} 
	\caption{The total channel gain $\Prx^{\mathrm{planar}\textrm{-}N} / \Ptx$ with a planar array with $\sqrt{N} \times \sqrt{N}$ equally spaced antennas. Each antenna has area $A=(\lambda/4)^2$, the wavelength is $\lambda=0.1$\,m, and $d=25$\,m.}
	\label{figure_propagationloss} 
\end{figure}
 
From the above discussion, a natural question arises: \emph{Will the received power grow linearly with $N$ for practical array sizes, so that we can utilize the approximation in \eqref{eq:received_power_planar-approx}, or do we need to use the exact expression?}

To answer this question, Fig.~\ref{figure_propagationloss} shows the total channel gain $\Prx^{\mathrm{planar}\textrm{-}N} / \Ptx \in [0,1]$ as a function of $N$, using either the exact expression in \eqref{eq:received_power_planar} or the far-field approximation in \eqref{eq:received_power_planar-approx}. We consider a setup with $d=25$\,m, $A=(\lambda/4)^2$, and $\lambda=0.1$\,m (corresponding to $f=3$\,GHz). The results of Fig.~\ref{figure_propagationloss} show that $10^5$ antennas are needed before the far-field approximation error is noticeable (above 5\%), and $10^8$ antennas are needed to approach the upper limit of $1/3$.

As a rule-of-thumb, the far-field approximation in \eqref{eq:received_power_planar-approx} is accurate for all $N$ satisfying $9 N A  \leq  d^2$ or, equivalently, satisfying
\begin{equation} \label{eq:rule-of-thumb}
3 \sqrt{NA} \leq d.
\end{equation}
The value $N=d^2/(9A)$ that gives equality in this rule-of-thumb is indicated by a circle in Fig.~\ref{figure_propagationloss}. The interpretation is that the far-field approximation is accurate as long as the distance $d$ to the transmitter is three times the width/height $\sqrt{NA}$ of the array. Hence, if $d=25$\,m, then the approximation can be applied for arrays up to $8.3 \times 8.3$\,m. 
Since $3 \approx 2 \sqrt{2}$, another way to phrase the rule-of-thumb is that the distance $d$ should be at least twice as long as the diagonal $\sqrt{2NA}$ of the array, which is the largest array dimension.
In any case, as the distance $d$ increases or the carrier frequency increases, the maximum number of antennas that satisfies the rule-of-thumb grows quadratically, but the area remains constant. In conclusion, the far-field approximation is usually accurate and might be used to predict scaling behaviors, but the exact expression in Corollary~\ref{cor:alpha-expression} is needed to study the asymptotic limit.

\begin{figure}[t!]
	\centering 
	\begin{overpic}[width=1.07\columnwidth,tics=10]{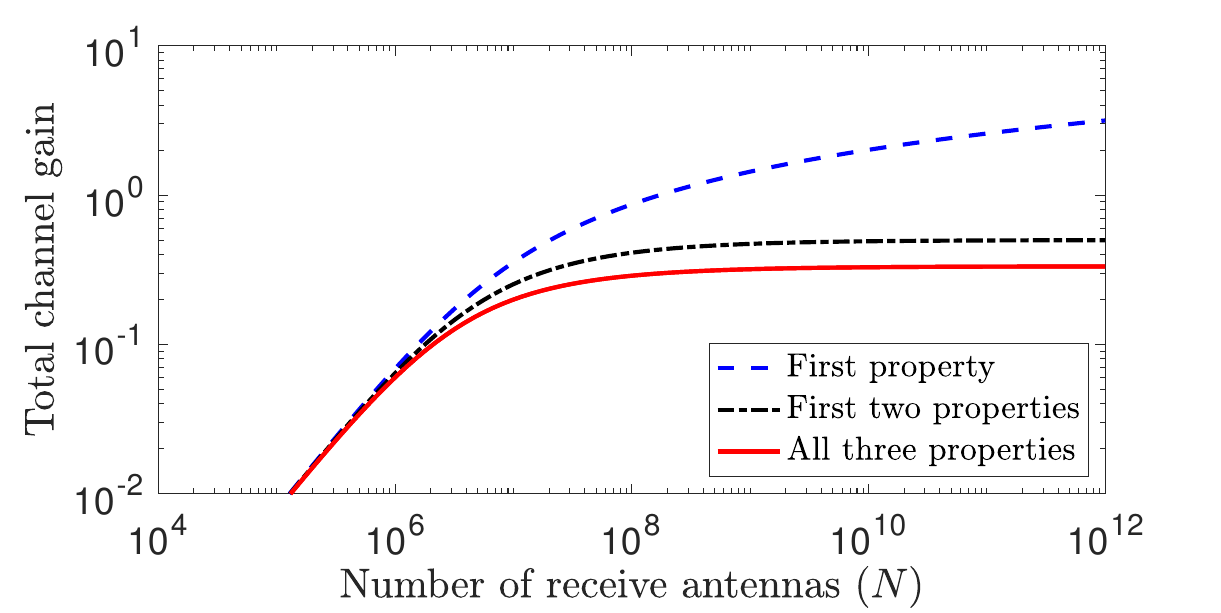}
\end{overpic} 
	\caption{The total channel gain $\Prx^{\mathrm{planar}\textrm{-}N} / \Ptx$ with a planar array with a varying number of antennas. 
	The exact curve (``All three properties'') is compared with approximations that are obtained by neglecting some of the near-field propagation properties.}
	\label{figure_NearFieldModeling} 
\end{figure}

Several recent works have also analyzed the propagation effects in the array's near-field \cite{Tang2019a,Garcia2019a,Ellingson2019a,Hu2018a,Bjornson2019f} but using less detailed models. Recall that three near-field properties were listed earlier in this section. The models in \cite{Tang2019a,Garcia2019a,Ellingson2019a} only capture the first property: that the distances to the antennas are different in large arrays. The models in \cite{Hu2018a,Bjornson2019f} also capture the second property: that the effective antenna areas vary over the array. A main novelty of this paper is that we also include the third property: variations in the polarization mismatch over the array.
The importance of considering all three properties when studying the near-field and asymptotic limits is emphasized in Fig.~\ref{figure_NearFieldModeling}. The figure considers the same setup as in Fig.~\ref{figure_propagationloss} but focuses on the upper tail where the near-field behavior occurs. If the polarization effects are neglected (``First two properties''), then the channel gain converges to $1/2$  as $N\to \infty$, as previously shown in \cite{Hu2018a,Bjornson2019f}. The general near-field behaviors are correct but not channel gain values.
If also the variations in effective areas are neglected (``First property''), then the channel gain diverges as $N\to \infty$ and thereby breaks the law of conservation of energy. Hence, the models from \cite{Tang2019a,Garcia2019a,Ellingson2019a} are not recommended to use when studying the asymptotic limits (or the near-field in general). However, all the models are accurate in the far-field.

\begin{remark}
In this paper, we define the far-field as being the propagation range between the array and the single-antenna transmitter where the far-field approximation gives accurate results. When the transmit antenna is isotropic, the border between the near-field and far-field is frequency-independent and proportional to the width/height $\sqrt{NA}$ of the receiving array, as seen from \eqref{eq:rule-of-thumb}. Note that the far-field definition for an array is conceptually different from the Fraunhofer distance, which is a frequency-dependent limit for the radiative far-field from a single antenna element. It describes the range at which one can neglect the reactive electromagnetic phenomena that appear within few wavelengths from an antenna. When considering large arrays in this paper, we will be in the far-field of the individual elements but in the near-field of the array.
\end{remark}

\begin{remark} \label{remark:generality}
The propagation models presented in this section are physically accurate, under the given assumptions, and will be used in the remainder of this paper. However, this does not mean that the assumptions are applicable in any conceivable practical setup---no model is generally applicable. For example, there can be other antenna gains, other polarization directions, and channels consisting of multiple paths. We are not focusing on these generalizations since we aim to provide an intuitive exposition of the fundamental behaviors. 
However, the generalization of the results are discussed in Section~\ref{sec:extensions}. 
\end{remark}

\section{Three Different MIMO Setups}
\label{sec:three-setups}

\begin{figure}[t!]
        \centering 
        \begin{subfigure}[b]{\columnwidth} \centering 
	\begin{overpic}[width=.98\columnwidth,tics=10]{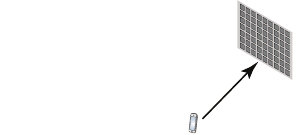}
	\put(68,0.5){\small Source}
	\put(47.5,38){\small mMIMO receiver}
	\put(47.5,33){\small with $N$ antennas}
	\put(80,11.5){\small $\vect{h}$}
\end{overpic} 
                \caption{The conventional uplink mMIMO setup.} \vspace{3mm}
                \label{fig:mMIMOexample}
        \end{subfigure} \hspace{3mm}
        \begin{subfigure}[b]{\columnwidth} \centering  \vspace{+2mm}
	\begin{overpic}[width=.98\columnwidth,tics=10]{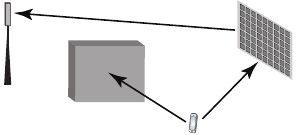}
	\put(-2,13){\small Destination}
	\put(68,0.5){\small Source}
	\put(22.5,7.5){\small Blocking object}
	\put(58,30){\small Relay with }
	\put(58,25){\small $N$ antennas}
	\put(39,41){\small $\vect{g}$} 
	\put(80,11.5){\small $\vect{h}$} 
\end{overpic}  
                \caption{The half-duplex mMIMO relay setup.}   \vspace{3mm}
                \label{fig:relayingexample} 
        \end{subfigure} 
        \begin{subfigure}[b]{\columnwidth} \centering  \vspace{+2mm}
	\begin{overpic}[width=.98\columnwidth,tics=10]{Figures/IRSexample_UL}
	\put(-2,13){\small Destination}
	\put(68,0.5){\small Source}
	\put(22.5,7.5){\small Blocking object}
	\put(62,30){\small IRS with }
	\put(62,25){\small $N$ elements}
	\put(39,41){\small $\vect{g}$} 
	\put(80,11.5){\small $\vect{h}$} 
\end{overpic}  
                \caption{The IRS-aided communication setup.\vspace{-0.1cm}}  
                \label{fig:IRSexample} 
        \end{subfigure} 
        \caption{Illustration of the three different MIMO setups compared in this paper.} 
        \label{fig:examples} 
\end{figure}
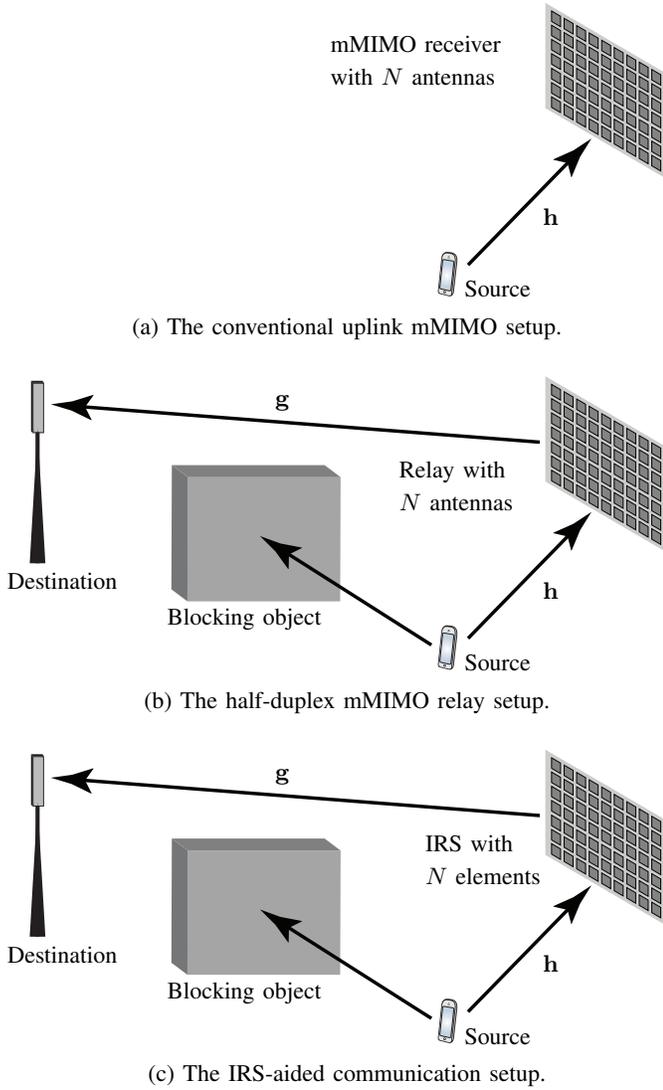

Next, we introduce the three different setups that are analyzed and compared in this paper, which are all illustrated in Fig.~\ref{fig:examples}. In the conventional mMIMO setup of Fig.~\ref{fig:mMIMOexample}, a single-antenna source transmits a signal that is received by a planar array with $N$ antennas, in the same form as in Fig.~\ref{fig:sphere3}. In the half-duplex mMIMO relay setup shown in Fig.~\ref{fig:relayingexample}, the same planar array receives the signal from the source and retransmits it to a single-antenna destination. In the IRS-aided setup in Fig.~\ref{fig:IRSexample}, the planar array is replaced by an IRS with $N$ passive elements that operate as a full-duplex relay that ``reflects'' the incoming signal in a controllable manner. The signal comes from the source and is supposed to reach the destination. The IRS is intelligent in the sense that each of the $N$ reflecting elements can control the individual phase of its diffusely reflected signal.  

Line-of-sight (LoS) propagation is considered in all setups. Since the channels are deterministic and thus can be estimated arbitrarily well from pilot signals, perfect channel state information is assumed. Despite simple, the three setups in Fig.~\ref{fig:examples} are sufficient to develop the fundamental scaling laws and near-field behaviors (see Remark~\ref{remark:generality}) and compare the setups.

\subsection{Conventional Uplink mMIMO}
In the LoS scenario, the deterministic flat-fading channel is represented by the vector $\vect{h} = [h_1,\ldots,h_N]^{\Ttran} \in \mathbb{C}^N$, where $h_n=|h_n| e^{-j\phi_n}$ is the channel from the source to the $n$th receive antenna with $|h_n|^2 \in [0,1]$ being the channel gain and $\phi_n \in [0,2\pi]$ an arbitrary phase shift. 
In the uplink, the received signal $\vect{r}_{\mathrm{mMIMO}} \in \mathbb{C}^N$ is  
\begin{equation}
\vect{r}_{\mathrm{mMIMO}}  = \vect{h} \sqrt{\Ptx} s +  \vect{n}  
\end{equation}
where $\Ptx$ is the transmit power, $s$ is the unit-norm information signal, and $\vect{n} \sim \CN(\vect{0},\sigma^2 \vect{I}_N)$ is the independent receiver noise. Under the assumption of perfect channel knowledge, linear receiver processing is optimal \cite{Telatar1999a,massivemimobook} and we let $\vect{v} \in \mathbb{C}^N$ denote the receive combining vector. 
It is well-known that the maximum SNR is achieved with maximum ratio (MR) combining, defined as $\vect{v} = \vect{h}^*/\|\vect{h} \|$ \cite{massivemimobook}. The SE is
\begin{equation} \label{eq:SE-mMIMO}
\log_2(1+\mathrm{SNR}_{\mathrm{mMIMO}} )
\end{equation}
with 
\begin{equation} \label{eq:SNR-mMIMO}
\mathrm{SNR}_{\mathrm{mMIMO}} =  \frac{|\vect{v}^{\Ttran}\vect{h}|^2}{ \| \vect{v}\|^2} \frac{ \Ptx}{\sigma^2} = \|\vect{h} \|^2 \frac{ \Ptx}{\sigma^2}  =  \left(\sum_{n=1}^{N}|h_n|^2\right)\frac{ \Ptx}{\sigma^2}.
\end{equation}

\subsection{Half-Duplex mMIMO Relay}

The half-duplex relay transmission takes place over two phases: 1) transmission from the source to the relay; 2) transmission from the relay to the destination. No direct link is present. Among the different relaying protocols (e.g.,\cite{Laneman2004a,Farhadi2009a,Khormuji2009a} among others), we consider the basic repetition-coded decode-and-forward protocol where equal time is allocated to the two phases. The first phase achieves the same SNR as in the mMIMO setup considered above. Therefore, the SE is $\frac{1}{2}\log_2(1+\mathrm{SNR}_{\mathrm{mMIMO}} )$ where $\mathrm{SNR}_{\mathrm{mMIMO}} $ is given in \eqref{eq:SNR-mMIMO} and the pre-log factor represents the fact that each phase is allocated half of the time resources. In the second phase, the relay retransmits the signal $s$ with power $\Prelay$ using a unit-norm precoding vector $\vect{w}$. The LoS channel from the array to the destination is represented by the deterministic vector $\vect{g} = [g_1,\ldots, g_N]^{\Ttran} \in \mathbb{C}^N$, where $g_n = |g_n|e^{-j\psi_n}$ represents the channel from the $n$th antenna to the receiver.  The received signal $r_\mathrm{relay} \in \mathbb{C}$ at the single-antenna destination is 
\begin{equation}
r_\mathrm{relay} = \vect{g}^{\Ttran} \vect{w} \sqrt{\Prelay} s + n
\end{equation}
where 
 $n \sim \CN(0,\sigma^2)$ is the independent receiver noise. 
It is well-known that the SNR is maximized by MR precoding with $\vect{w} = \vect{g}^*/\|\vect{g} \|$ \cite{massivemimobook}, which leads to
\begin{equation}
\mathrm{SNR}_{\mathrm{relay}} = \|\vect{g} \|^2 \frac{ \Prelay}{ \sigma^2}  =\left(\sum_{n=1}^{N}|{g_n}|^2\right) \frac{\Prelay}{\sigma^2} .
\end{equation}
The SE of the end-to-end mMIMO relay channel is then given by the minimum of the two phases:
\begin{equation} \label{eq:SE-relaying}
\mathrm{SE}_{\mathrm{relay}} = \frac{1}{2} \log_2 \big( 1 + \min\left(\mathrm{SNR}_{\mathrm{mMIMO}}, \mathrm{SNR}_{\mathrm{relay}} \right) \big).
\end{equation}

\subsection{IRS-aided Communication}

The IRS-aided communication is also a relaying setup, thus the system model resembles the half-duplex mMIMO relay case with the key differences that each element in the IRS scatterers the incoming signal with a controllable phase-shift but without increasing its power or requiring a separate retransmission phase. 
The received signal $r_{\mathrm{IRS}} \in \mathbb{C}$ can be modeled as \cite{Ozdogan2019a,Ellingson2019a} 
\begin{equation}
r_{\mathrm{IRS}}  = \vect{g}^{\Ttran} \vect{\Theta} \vect{h} \sqrt{\Ptx} s + n
\end{equation}
where $\Ptx$ and $s$ are the same as in the previous setups and $n \sim \CN(0,\sigma^2 )$ is the noise at the receiver. The reflection properties are determined by the diagonal matrix
\begin{equation}
\vect{\Theta} =  \diag\left( \mu_1 e^{j \theta_1}, \ldots, \mu_N e^{j \theta_N} \right)
\end{equation}
where $\mu_1,\ldots,\mu_N \in [0,1]$ are the amplitude scattering variables (describing the fraction of the incident signal power that is scattered) and $\theta_1,\ldots,\theta_N \in [0,2\pi)$ are the phase-shift variables (describing the delays of the scattered signals). These parameters can be optimized based on $\vect{g}$ and $\vect{h}$. With perfect channel knowledge \cite{Wu2018a,Bjornson2019e}, an achievable SE is\footnote{Note that we are considering relaying operation where the IRS is unaware of the information. A higher SE can be achieved by also encoding information into the matrix $\vect{\Theta}$ \cite{Karasik2019a}.} 
\begin{equation}
\log_2(1+\mathrm{SNR}_{\mathrm{IRS}} )
\end{equation}
where
\begin{align} \notag
\mathrm{SNR}_{\mathrm{IRS}} &=  | \vect{g}^{\Ttran} \vect{\Theta} \vect{h} |^2 \frac{  \Ptx}{\sigma^2} \\ &=  \left| \sum_{n=1}^{N} \mu_n |h_n||g_n| e^{j(\theta_n-\phi_n-\psi_n)} \right|^2\frac{  \Ptx}{\sigma^2}
\label{eq:SNR:IRS1}
\end{align}
is the SNR at the receiver. We will optimize the amplitude and phase-shift variables in the next section.

\section{Power Scaling Laws and Near-field Behaviors}
\label{sec:power-scaling-laws}

We will now investigate the asymptotic behaviors of the three setups defined in Section~\ref{sec:three-setups}. Particularly, the power scaling laws, near-field behaviors, and asymptotic SE limits will be analyzed as $N$ increases. New insights into the fundamental properties will be obtained by utilizing the deterministic propagation model derived in Section~\ref{planar_arrays}. The following assumption is made for all setups.

\begin{figure}[t!]
        \centering 
        \begin{subfigure}[b]{\columnwidth} \hfill
	\begin{overpic}[width=.9\columnwidth,tics=10]{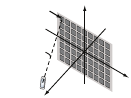}
	\put(12,10){$(x_t,y_t,d)$}
	\put(18,72){Antenna $n$:}
	\put(18,67){$(x_n,y_n,0)$}
	\put(10,41){Distance}
	\put(-11,36){\small $\sqrt{(x_n\!-\!x_t)^2\!+\!(y_n\!-\!y_t)^2\!+\!d^2}$}
	\put(96.5,19){$X$}
	\put(68,66){$Y$}
	\put(39,0){$Z$}
\end{overpic} 
                \caption{A source at an arbitrary location $(x_t,y_t,d)$ transmits to an planar array located in the $XY$-plane.} \vspace{3mm}
                \label{figure_geometricsetup}
        \end{subfigure} 
        \begin{subfigure}[b]{\columnwidth} \centering  \vspace{+2mm}
	\begin{overpic}[width=.86\columnwidth,tics=10]{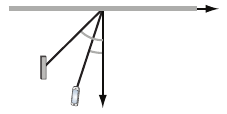}
	\put(40,50){Array}
	\put(32,20){$d$}
	\put(27,32){$\delta$}
	\put(35,30){$\omega$}
	\put(41,23){$\eta$}
	\put(48,1){$Z$}
	\put(95,40){$X$}
	\put(20,-1){Source}
	\put(1,10){Destination}
\end{overpic} 
                \caption{In the analytical parts, the source is at distance $d$ in angle $\eta$, while the destination is at distance $\delta$ in angle $\omega$.}  
                \label{figure_figure_from_above} 
        \end{subfigure} 
        \caption{Geometric illustration of the setup defined by Assumptions~\ref{assumption1} and \ref{assumption2} that is used for analyzing the SNR behavior in the near- and far-fields.} 
        \label{fig:geometric-illustration} 
\end{figure}
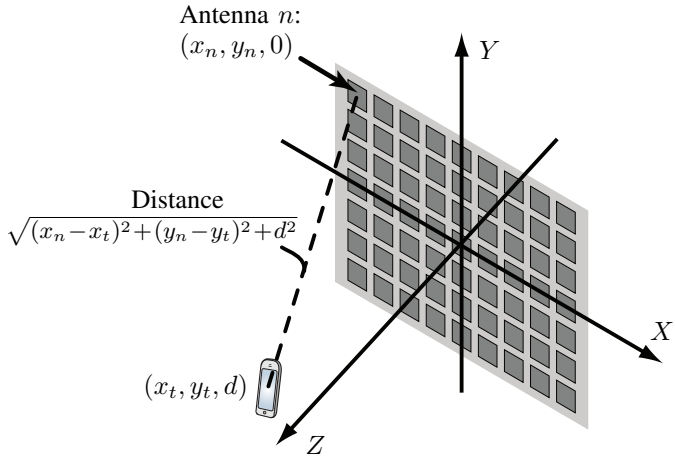
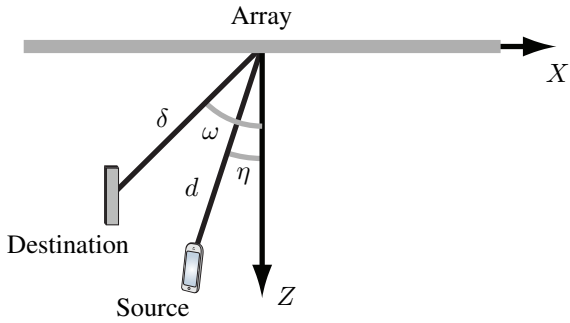

\begin{assumption} \label{assumption2}
The planar array is centered around the origin in the $XY$-plane, as illustrated in Fig.~\ref{figure_geometricsetup}. The source
is located in the $XZ$-plane at distance $d$ from the center of the array with angle $\eta \in [-\pi/2,\pi/2]$, as illustrated in Fig.~\ref{figure_figure_from_above}. It sends a signal that has polarization in the $Y$ direction when traveling in the $Z$ direction.
\end{assumption}

Under Assumption~\ref{assumption2}, the source is located at $\vect{p}_t= (d \sin(\eta), 0, d \cos(\eta))$ and the $n$th antenna is centered at $\vect{p}_n= (x_n,y_n,0)$. If we number the antennas from left to right, row by row, according to Fig.~\ref{figure_geometricsetup}, the coordinates $x_n$ and $y_n$ of the $n$th receive antenna for $n=1,\ldots,N$ are
\begin{align} \label{eq:xn}
x_n &= - \frac{(\sqrt{N}-1)\sqrt{A}}{2} + \sqrt{A} \, \mod(n-1,\sqrt{N}) \\
y_n &= \frac{(\sqrt{N}-1)\sqrt{A}}{2} - \sqrt{A} \left\lfloor \frac{n-1}{\sqrt{N}} \right\rfloor. \label{eq:yn}
\end{align}

\subsection{Conventional Uplink mMIMO}

We will now study the mMIMO setup in detail.
Following the geometry stated in Assumption~\ref{assumption2}, we have that $\vect{p}_t=(d \sin(\eta),0,d \cos(\eta))$ and $\vect{p}_n=(x_n,y_n,0)$ where the coordinates $x_n$ and $y_n$ are defined in \eqref{eq:xn} and \eqref{eq:yn}. By using Lemma~\ref{lemma1}, the channel $h_n=|h_n| e^{-j\phi_n}$ to the $n$th receive antenna is obtained as
\begin{align} \label{eq:hn2}
|h_n|^2 &= \zeta_{(d \sin(\eta),0,d \cos(\eta)),(x_n,y_n,0),\sqrt{A}}
\end{align}
with the phase computed based on the propagation delay as
\begin{align} \notag
\phi_n &= 2\pi \cdot\mod \left( \frac{{||\vect{p}_t- \vect{p}_n||}}{\lambda}  ,1 \right) \\  \notag &= 2\pi \cdot\mod \left( \frac{\sqrt{(x_n-d\sin(\eta))^2+y_n^2+d^2 \cos^2(\eta)}}{\lambda}  ,1 \right)\\&
= 2\pi \cdot\mod \left( \frac{\sqrt{x_n^2+y_n^2+d^2 - 2d x_n \sin(\eta)}}{\lambda}  ,1 \right).
\end{align}
The following result is then obtained.

\begin{proposition} \label{prop:SNR-mMIMO2}
Under Assumptions~\ref{assumption1} and \ref{assumption2}, in the mMIMO setup, the SNR with MR combining becomes
\begin{equation} \label{eq:SNR_mMIMO_general}
\mathrm{SNR}_{\mathrm{mMIMO}}  =  \xi_{d,\eta,N}\frac{ \Ptx}{\sigma^2} 
\end{equation}
where the total channel gain $\xi_{d,\eta,N} $ is given by
\begin{align} \notag
&\xi_{d,\eta,N} \\ \notag &
= \sum_{i=1}^{2} \Bigg( \frac{ B +(-1)^i \sqrt{B} \tan(\eta)  }{6 \pi (B+1)\sqrt{ 2B+\tan^2(\eta) + 1 + 2(-1)^i \sqrt{B} \tan(\eta)} } \\ &+ \frac{1}{3\pi} \tan^{-1} \Bigg(  \frac{ 
B +(-1)^i \sqrt{B} \tan(\eta) 
  }{
  \sqrt{2B+\tan^2(\eta) + 1 + 2(-1)^i \sqrt{B} \tan(\eta)}
  }
\Bigg) \Bigg) \label{eq:xi-mMIMO}
\end{align}
with $B = N \pi \beta_{d \cos(\eta)} = \frac{NA}{4 d^2 \cos^2(\eta)}$.

\end{proposition}
\begin{IEEEproof}
This result follows from Lemma~\ref{lemma1} with $\vect{p}_t=(d \sin(\eta), 0, d \cos(\eta))$, $\vect{p}_n=(0,0,0)$, and $a = \sqrt{NA}$.
\end{IEEEproof}

We stress that the channel gain in \eqref{eq:xi-mMIMO} depends only on the total array area $NA$ (see Remark~\ref{remark:NA-scaling}), thus the choice of frequency band only affects how many antennas are needed to achieve that area.
By using Corollary~\ref{cor:alpha-expression}, a more compact expression can be obtained when the transmitter is centered in front of the array (i.e., $\eta=0$).

\begin{corollary} \label{cor:SNR-mMIMO}
When the transmitter is located in direction $\eta=0$, the SNR in \eqref{eq:SNR_mMIMO_general} simplifies to 
\begin{equation} \label{eq:SNR-mMIMO_planar}
\mathrm{SNR}_{\mathrm{mMIMO}} = \alpha_{d,N}\frac{\Ptx}{\sigma^2}
\end{equation}
where the total channel gain $\alpha_{d,N}$ is given in \eqref{eq:alpha-expression}.
\end{corollary}

We will now use the general expression in Proposition~\ref{prop:SNR-mMIMO2} for an arbitrary $\eta$ to study the far-field behavior in the next corollary. Note that $d \cos(\eta)$ is the distance from the transmitter to the plane where the array is deployed and $\sqrt{N A}$ is the width/height of the array.

\begin{corollary}[Far-field approximation]  \label{cor:far-field-mMIMO}
If the transmitter is in the far-field of the mMIMO receiver, in the sense that $d \cos(\eta) \gg \sqrt{N A}$, then \eqref{eq:SNR-mMIMO_planar} is well approximated as
\begin{equation} \label{eq:SNR-mMIMO_planar-farfield}
\mathrm{SNR}_{\mathrm{mMIMO}} \approx \mathrm{SNR}_{\mathrm{mMIMO}}^{\mathrm{ff}}  = N  \varsigma_{d,\eta} \frac{ \Ptx}{\sigma^2} 
\end{equation}
where 
\begin{equation} \label{eq:updated-beta}
\varsigma_{d,\eta} =
\beta_{d \cos(\eta)} \cos^3(\eta)
\end{equation}
and $\beta_{d \cos(\eta)}$ is given in \eqref{eq:beta-definition}.
\end{corollary}
\begin{IEEEproof}
The derivation can be found in Appendix~\ref{app:proof-cor:far-field-mMIMO}.
\end{IEEEproof}

Plugging \eqref{eq:beta-definition} into \eqref{eq:updated-beta}, we have that \eqref{eq:SNR-mMIMO_planar-farfield} reduces to
\begin{align} \notag
\mathrm{SNR}_{\mathrm{mMIMO}}^{\mathrm{ff}}  &= N  \frac{A}{4\pi \big(d\cos(\eta)\big)^2} \cos^3(\eta) \frac{ \Ptx}{\sigma^2} \\& =N\frac{ \Ptx}{\sigma^2}\overbrace{A\cos(\eta)}^{{\text{Effective area}}} \hspace{-0.7cm}\underbrace{\frac{1}{4\pi d^2}}_{{\text{Free-space propagation}}} \hspace{-0.5cm}\label{eq:SNR-mMIMO_planar-farfield_simplified}
\end{align}
which is equivalent to \cite[Eq. (17)]{bjornson2019Asilomar}. This shows that, in the far-field, the channel gain per antenna is computed according to Friis' formula with the effective antenna area being $A\cos(\eta)$ \cite{friis1946note}. This is a consequence of the fact that, although each antenna has a physical size of $\sqrt{A}\times \sqrt {A}$, its effective size shrinks to $\sqrt{A} \cos(\eta)\times \sqrt {A}$ when observed from the direction of the transmitter.

From Corollary~\ref{cor:far-field-mMIMO}, we notice that the far-field SNR in \eqref{eq:SNR-mMIMO_planar-farfield} is proportional to $N$, which is consistent with previous work in the mMIMO literature \cite{Ngo2013a,Hoydis2013a,massivemimobook}. Hence, when  $N$ increases, the system can either benefit from a linearly increasing SNR or reduce $\Ptx$ as $1/N$ to keep the SNR constant. The latter is the conventional power scaling law for mMIMO, which first appeared in \cite{Ngo2013a,Hoydis2013a}. However, when computing the asymptotic behavior as $N \to \infty$, these prior works implicitly assumed the transmitter remains in far-field of the array and thus that the SNR goes to infinity as $N \to \infty$ (or the power can be brought down to zero following the scaling law, while the SNR remains strictly non-zero). This is not physically possible. 
As $N$ increases, the far-field approximation eventually breaks down and the total channel gain saturates in the near-field, as illustrated in Fig.~\ref{figure_propagationloss}. We provide the following novel asymptotic limit and power scaling characterization for the mMIMO receiver.

\begin{corollary}[Asymptotic analysis]  \label{cor:limit-mMIMO}
As $N \to \infty$ with a constant transmit power $\Ptx$, the SNR with MR combining satisfies
\begin{equation} \label{eq:limit-mMIMO}
\mathrm{SNR}_{\mathrm{mMIMO}} \to \frac{1}{3}\frac{\Ptx}{ \sigma^2}.
\end{equation}
If the transmit power is reduced with $N$ as $\Ptx = P/N^{\rho}$ for some constant $P>0$ and exponent $\rho>0$, then as $N \to \infty$
\begin{equation} \label{eq:limit-mMIMO2}
\mathrm{SNR}_{\mathrm{mMIMO}} = \xi_{d,\eta,N} \frac{ P}{\sigma^2 N^{\rho}} \to 0.
\end{equation}
\end{corollary}
\begin{IEEEproof}
The limit in \eqref{eq:limit-mMIMO} is computed in the same way as the finite limit in \eqref{eq:limit-planar-array}. Since $P \xi_{d,\eta,N} $ has a finite limit and $1/N^{\rho}\to 0$ as $N \to \infty$, the result in \eqref{eq:limit-mMIMO2} follows directly.
\end{IEEEproof}

\begin{figure}[t!]
	\centering 
	\begin{overpic}[width=1.07\columnwidth,tics=10]{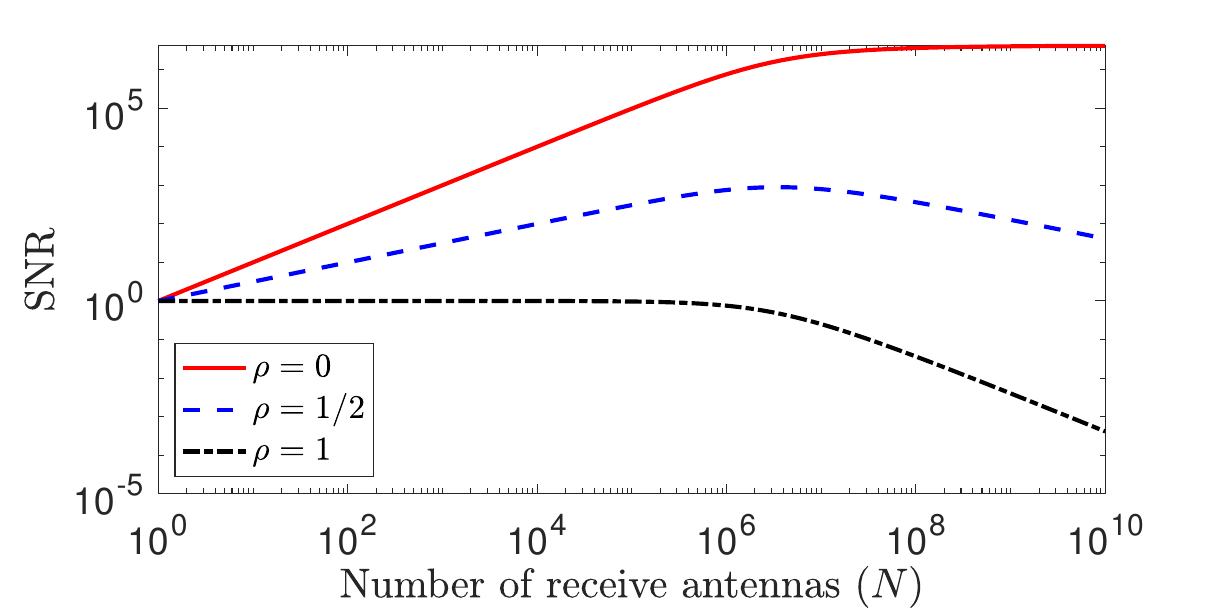}
\end{overpic} 
	\caption{The SNR value $\mathrm{SNR}_{\mathrm{mMIMO}} $ in \eqref{eq:SNR-mMIMO_planar} when scaling down the transmit power as $\Ptx = P/N^{\rho}$ for $\rho \in \{0, 1/2, 1\}$. The setup is given by $d=25$\,m, $\eta=0$, $A=(\lambda/4)^2$, $\lambda=0.1$\,m, and $P/\sigma^2$ is selected to give $\mathrm{SNR}_{\mathrm{mMIMO}}=0$\,dB for $N=1$.}
	\label{figure_simulationScalingLaw_mMIMO}  
\end{figure}

This corollary shows that any power scaling for which $\Ptx \to 0$ as $N \to \infty$ will  asymptotically lead to zero SNR. Hence, the asymptotic motivation behind the power scaling laws in the mMIMO literature \cite{Ngo2013a,Hoydis2013a} is not correct. The scaling laws are, nevertheless, useful in many practical situations. To demonstrate this, Fig.~\ref{figure_simulationScalingLaw_mMIMO} shows $\mathrm{SNR}_{\mathrm{mMIMO}} $ in \eqref{eq:SNR-mMIMO_planar} when we scale down the transmit power as $\Ptx = P/N^{\rho}$ for $\rho \in \{0, 1/2, 1\}$, where $\rho=0$ corresponds to constant power. We consider a setup where the transmitter location is given by $d=25$\,m and $\eta=0$, the antenna area is $A=(\lambda/4)^2$, the wavelength is $\lambda=0.1$\,m, and the transmit power is selected so that $P \xi_{d,\eta,N}/\sigma^2=0$\,dB for $N=1$.
We observe that for $\rho=0$, the far-field behavior, namely, an SNR that grows linearly with $N$, approximately holds true for any $N \le 10^6$. This probably includes all cases of practical interest since the array can be up to $25 \times 25$\,m. If one selects $\rho=1/2$, the SNR will instead grow as $\sqrt{N}$ for $N \le 10^6$. Moreover, for $\rho=1$, the SNR is approximately constant for $N \le 10^6$. For larger values of $N$, the SNR goes to zero whenever $\rho>0$, as proved in Corollary~\ref{cor:limit-mMIMO}.

Since this example considers $\eta=0$, we know from Corollary~\ref{cor:SNR-mMIMO} that $\xi_{d,0,N} =  \alpha_{d,N}$. It is the relation between $N$ and $d$ in $\alpha_{d,N}$ that determines when the far-field behavior breaks down. 
Since these variables enter into $\alpha_{d,N}$ as $N \beta_d=\frac{NA}{4\pi d^2}$, the far-field behavior appears as long as $N/d^2 \leq 10^6/25^2 = 1600$. Hence, even if we would reduce the propagation distance to $d=2.5$\,m, the approximate scaling laws will be accurate for $N\leq 10^4$ or arrays being up to $2.5 \times 2.5$\,m.
In conclusion, the conventional power scaling laws \emph{can be safely applied in many practical scenarios}, but if we truly want to let $N \to \infty$ or study the case where the transmitter is very close to the array, the asymptotically accurate behavior derived in Corollary~\ref{cor:limit-mMIMO} must be considered. 

\subsection{Half-Duplex mMIMO Relay}

We now turn the attention to the half-duplex mMIMO relay setup. We assume the destination is equipped with a lossless isotropic antenna located in the $XZ$-plane at distance $\delta$ from the center of the array with angle $\omega \in [-\pi/2,\pi/2]$, as shown in Fig.~\ref{figure_figure_from_above}. This means that it is located at $(\delta \sin(\omega), 0, \delta \cos(\omega))$. According to the geometry stated in Assumption~\ref{assumption2}, the destination is located at $(\delta \sin(\omega), 0, \delta \cos(\omega))$ and the $n$th transmit antenna at $(x_n,y_n,0)$, where $x_n$ and $y_n$ are defined in \eqref{eq:xn} and \eqref{eq:yn}, respectively.
{We assume each antenna is sufficiently small to radiate the signal isotropically into the half-plane in front of it.\footnote{This property holds for antennas with $a\leq \lambda/4$ for essentially the same reason as the approximation in Lemma~\ref{lemma1} being tight in this interval. An illustration of the radiation pattern for $a=\lambda/5$ is found in \cite[Fig.~2]{Ozdogan2019a}.}
 From Lemma~\ref{lemma1}, it follows that the channel $g_n=|g_n| e^{-j\psi_n}$ from the $n$th antenna to the destination is given by}
\begin{align}  \label{eq:gn2}
|g_n|^2&= \zeta_{(\delta \sin(\omega), 0, \delta \cos(\omega)),(x_n,y_n,0),\sqrt{A}}
\end{align}
where the propagation delay implies that
\begin{align}
\psi_n 
&=  2\pi \cdot\mod \left( \frac{\sqrt{x_n^2 +y_n^2+\delta^2 - 2\delta x_n \sin(\omega) }}{\lambda}  ,1 \right).
\end{align}
The following result is then obtained.

\begin{proposition} \label{proposition:relaying-phase2}
Suppose the destination is located at distance $\delta$ in angle $\omega \in [-\pi/2,\pi/2]$ from the center of the mMIMO relay. Under Assumptions~\ref{assumption1} and \ref{assumption2}, the SNR with MR precoding becomes
\begin{equation}
\mathrm{SNR}_{\mathrm{relay}}  = \xi_{\delta,\omega,N}\frac{ \Prelay}{\sigma^2}
\end{equation}
where $\xi_{\delta,\omega,N} $ is given in \eqref{eq:xi-mMIMO}
with 
 $B = N \pi \beta_{\delta \cos(\omega)} = \frac{NA}{4 \delta^2 \cos^2(\omega)}$.
\end{proposition}
\begin{IEEEproof}
This result follows from Lemma~\ref{lemma1} with $\vect{p}_t=(\delta \sin(\omega), 0, \delta \cos(\omega))$, $\vect{p}_n=(0,0,0)$, and $a = \sqrt{NA}$.
\end{IEEEproof}

By utilizing the results in Proposition~\ref{prop:SNR-mMIMO2} and Proposition~\ref{proposition:relaying-phase2}, the end-to-end SE in \eqref{eq:SE-relaying} can be rewritten as
\begin{equation} \label{eq:SE-relaying2}
\mathrm{SE}_{\mathrm{relay}} = \frac{1}{2} \log_2 \left( 1 + \frac{\min(\xi_{d,\eta,N} \Ptx, \xi_{\delta,\omega,N} \Prelay )}{\sigma^2}   \right).
\end{equation}
Note that, when the receiver is centered in front of the array (i.e., $\omega=0$), we have $\xi_{\delta,0,N}=\alpha_{\delta,N}$.
 
Just as in the uplink mMIMO setup, the SNR expression takes a simpler approximate form in the far-field, which is now jointly represented by  $d \cos(\eta) \gg \sqrt{N A}$ and $\delta \cos(\omega) \gg \sqrt{NA}$. Following a similar approach as in the proof of Corollary~\ref{cor:far-field-mMIMO}, we have that 
\begin{equation}
\mathrm{SNR}_{\mathrm{relay}} \approx N  \varsigma_{\delta,\omega}\frac{ \Prelay}{\sigma^2}
\end{equation}
in the far-field, where $\varsigma_{\delta,\omega}$ is defined as in \eqref{eq:updated-beta}.
In conjunction with the far-field result in Corollary~\ref{cor:far-field-mMIMO}, we obtain the following result in the mMIMO relay setup.

\begin{corollary} [Far-field approximation]  \label{cor:far-field-relay}
If the source and destination are both in the far-field of the mMIMO relay, in the sense that $d \cos(\eta) \gg \sqrt{N A}$ and $\delta \cos(\omega) \gg \sqrt{NA}$, then \eqref{eq:SE-relaying2} is well approximated as
\begin{equation} \label{eq:SE-relaying-farfield}
\mathrm{SE}_{\mathrm{relay}} \approx \frac{1}{2} \log_2 \left( 1 +N
 \frac{ \min \left( \Ptx  \varsigma_{d,\eta},  \Prelay  \varsigma_{\delta,\omega}  \right)}{\sigma^2}   \right).
\end{equation}
\end{corollary}

This corollary shows that the end-to-end SNR grows proportionally to $N$ whenever the far-field approximation is applicable. Hence, one can either keep the transmit powers fixed and achieve an SNR that grows proportionally to $N$, or reduce the transmit powers $\Ptx$ and $\Prelay$ as $1/N$ and achieve the same SNR as with $N=1$.
Since the half-duplex relay channel is a time-multiplexed composition of one uplink and one downlink mMIMO channel, the insights from the last subsection still apply: the far-field approximation and the power scaling law hold in most cases of practical interest. 
However, in the asymptotic limit as $N \to \infty$, we have that
$\xi_{\delta,\omega,N} \to \frac{1}{3}$ which is the same asymptotic limit as in the first phase where it holds that $\xi_{d,\eta,N} \to \frac{1}{3}$. The following corollary shows that the power scaling law breaks down asymptotically.

\begin{corollary}[Asymptotic analysis]  \label{cor:limit-relaying}
As $N \to \infty$ with constant transmit powers $\Ptx$ and $\Prelay$, the SE with the mMIMO relay satisfies
\begin{equation} \label{eq:SE-relaying-limit}
\mathrm{SE}_{\mathrm{relay}} \to \frac{1}{2} \log_2 \left( 1+ \frac{1}{3}\frac{\min(\Ptx,\Prelay)}{ \sigma^2} \right).
\end{equation}
If the transmit powers are reduced with $N$ as $\Ptx = P_1/N^{\rho_1}$ and  $\Prelay = P_2/N^{\rho_2}$  for some constants $P_1,P_2>0$ and exponents $\rho_1,\rho_2>0$, then as $N \to \infty$ it follows that 
\begin{equation}
\mathrm{SE}_{\mathrm{relay}} \to 0.
\end{equation}
\end{corollary}
\begin{IEEEproof}
The proof follows that of Corollary~\ref{cor:limit-mMIMO} and is therefore omitted.
\end{IEEEproof}

This scaling behavior is essentially the same as the one illustrated in Fig.~\ref{figure_simulationScalingLaw_mMIMO}, thus we postpone the numerical comparison with uplink mMIMO to Section~\ref{sec:how-many-elements}.
There are plenty of previous works that study mMIMO relays and the related power scaling laws \cite{Ngo2014c,Yang2015a,Kong2018a}, often in more general setups (e.g., full-duplex or two-way relaying) than those considered in this paper. Although the power scaling laws developed in those papers are practically relevant, the non-zero asymptotic limits are incorrect since the channel models that are used are asymptotically inaccurate. Since the total channel gain is upper bounded by one, any power scaling law that leads to zero transmit power as $N \to \infty$ must also have a zero-valued asymptotic SE. Corollary~\ref{cor:limit-relaying} demonstrates this in a simple decode-and-forward relay setup but the result naturally extends to more complicated setups.

\subsection{IRS-aided Communication}

We begin by observing that the SNR is maximized in \eqref{eq:SNR:IRS1} when all the terms in the summation has the same phase \cite{Wu2018a,Bjornson2019e}. This is achieved, for example, by selecting $\theta_n = \phi_n + \psi_n$ for $n=1,\ldots,N$. In this case, all the $N$ terms have a positive contribution to the SNR, which is maximized by setting $\mu_1 = \ldots = \mu_N = 1$ so that all terms take their maximum achievable value.\footnote{Depending on the IRS implementation, the phases and amplitudes might not be independently controllable but coupled \cite{Abeywickrama2020a}. Moreover, there are some implementations that allow for controlling the state of an element over a continuous range, while other implementations have a discrete number of possible states \cite{Tsilipakos2020a,Wu2020b}. To characterize the ultimate performance, we consider the ideal case when we can optimize the phases and amplitudes independently with arbitrary precision. A practical IRS might require more elements to deliver the same SNR as described in this paper.}
In doing this, \eqref{eq:SNR:IRS1} becomes
\begin{equation} \label{eq:SNR:IRS2}
\mathrm{SNR}_{\mathrm{IRS}} =   \frac{ \Ptx}{\sigma^2}
\left( \sum_{n=1}^{N} |{h_n}||{g_n}| \right)^2.
\end{equation}
We can compute this expression exactly using \eqref{eq:hn2} and \eqref{eq:gn2}, which provides the values of $|{h_n}|$ and $|{g_n}|$. We can also obtain the following simple upper bound that does not involve any summations and will shown to be tight.

\begin{proposition} \label{prop:IRS-upperbound}
The SNR in \eqref{eq:SNR:IRS2} with optimal phase-shifts can be upper bounded as
\begin{align} \notag
\mathrm{SNR}_{\mathrm{IRS}} \leq \mathrm{SNR}_{\mathrm{IRS}}^{\mathrm{upper}} &= \left( \sum_{n=1}^{N} |{h_n}|^2  \right) \!\! \left( \sum_{n=1}^{N} |{g_n}|^2  \right) \frac{ \Ptx}{\sigma^2}
 \\ \notag&
  = \xi_{d,\eta,N} \xi_{\delta,\omega,N} \frac{ \Ptx}{\sigma^2} \\ &= \xi_{\delta,\omega,N} \mathrm{SNR}_{\mathrm{mMIMO}}
\label{eq:SNR:IRS2-bound}
\end{align}
with $\mathrm{SNR}_{\mathrm{mMIMO}}$ given in \eqref{eq:SNR_mMIMO_general}. The equality holds if and only if the vectors $[|{h_1}| , \ldots, |{h_N}|]^{\Ttran}$ and  $[|{g_1}| , \ldots, |{g_N}|]^{\Ttran}$ are parallel.
\end{proposition}
\begin{IEEEproof}
The inequality is a direct application of H\"older's inequality, followed by computing the total channel gains of the two links using Proposition~\ref{prop:SNR-mMIMO2}. 
\end{IEEEproof}

Interestingly, the upper bound in \eqref{eq:SNR:IRS2-bound} is the product of the SNR in the mMIMO setup and the term $\xi_{\delta,\omega,N}$, which is the total channel gain from the IRS to the destination. Section~\ref{planar_arrays} described that the value of $\xi_{\delta,\omega,N}$ must be below one (or rather $1/3$) due to the law of conservation of energy. Therefore, Proposition~\ref{prop:IRS-upperbound} implicitly states that the IRS-aided setup cannot achieve a higher SNR than the corresponding mMIMO setup, if the array sizes are equal. One way to interpret this result is that the IRS acts as an uplink mMIMO receiver that uses the receive combining $\vect{v} = \vect{\Theta}^{\Ttran} \vect{g}$, which has a different directivity than the channel $\vect{h}$, except when $[|{h_1}| , \ldots, |{h_N}|]^{\Ttran}$ and  $[|{g_1}| , \ldots, |{g_N}|]^{\Ttran}$ are parallel vectors. Moreover, it also incurs an additional SNR loss given by
\begin{equation}
\| \vect{v} \|^2 = \|  \vect{\Theta}^{\Ttran} \vect{g} \|^2  =  \| \vect{g} \|^2   = \xi_{\delta,\omega,N} < 1.
\end{equation}
Similar conclusions hold when the IRS is compared to the half-duplex mMIMO relay. To see this, assume for simplicity that the transmit power is the same in the two phases (i.e., $\Prelay = \Ptx$). In this case, we may equivalently rewrite \eqref{eq:SNR:IRS2-bound} as
\begin{equation}
\mathrm{SNR}_{\mathrm{IRS}}^{\mathrm{upper}} =  \xi_{d,\eta,N} \mathrm{SNR}_{\mathrm{relay}}
\end{equation}
which can never be higher than $\mathrm{SNR}_{\mathrm{relay}}$, based on the same arguments as above.
Since the end-to-end SNR of the mMIMO relay channel is the minimum of the SNRs in the two phases, i.e., $\min(\mathrm{SNR}_{\mathrm{relay}},\mathrm{SNR}_{\mathrm{mMIMO}})$, and both are higher than $\mathrm{SNR}_{\mathrm{IRS}}^{\mathrm{upper}}$, we can conclude that the IRS can never achieve a higher SNR than the corresponding mMIMO relay setup with a matching array size and transmit power. However, the half-duplex relay suffers from the $1/2$ pre-log factor in \eqref{eq:SE-relaying}, which can potentially make the IRS more spectrally efficient, even if the SNR is lower. To investigate this further, assume that $\mathrm{SNR}_{\mathrm{relay}} > \mathrm{SNR}_{\mathrm{mMIMO}}$ so that \eqref{eq:SE-relaying} becomes
\begin{equation} \label{eq:SE-relaying-1}
\mathrm{SE}_{\mathrm{relay}} = \frac{1}{2} \log_2 \left( 1 + \mathrm{SNR}_{\mathrm{mMIMO}} \right).
\end{equation}
From \eqref{eq:SNR:IRS2-bound}, the SE with the IRS is upper bounded by $\log_2(1+ \xi_{\delta,\omega,N} \mathrm{SNR}_{\mathrm{mMIMO}} )$, which is higher than \eqref{eq:SE-relaying-1} when
\begin{equation}
 \xi_{\delta,\omega,N} > \frac{\sqrt{1 + \mathrm{SNR}_{\mathrm{mMIMO}}} - 1}{\mathrm{SNR}_{\mathrm{mMIMO}}}.
\end{equation}
This condition will be satisfied if $\mathrm{SNR}_{\mathrm{mMIMO}}$ is sufficiently large. Hence, there are high-SNR cases when the IRS-aided setup outperforms the mMIMO relay. This observation is in line with previous results in \cite{Huang2018a,Bjornson2019e}.

We will now study the power scaling law. Recall from \eqref{eq:SNR:IRS2} that the SNR is proportional to the square of a sum with $N$ terms. Intuitively, the SNR may then grow quadratically with $N$. 
That behavior can in fact be observed in the far-field.

\begin{corollary}[Far-field approximation] \label{cor:far-field-IRS}
If both the source and destination are in the far-field of the IRS, in the sense that $d\cos(\eta) \gg \sqrt{N A}$ and $\delta \cos(\omega) \gg \sqrt{NA}$, the SNR in \eqref{eq:SNR:IRS2} can be approximated as
\begin{equation} \label{eq:SNR:IRS2-far3}
\mathrm{SNR}_{\mathrm{IRS}} \approx \mathrm{SNR}_{\mathrm{IRS}}^{\mathrm{ff}} = N^2 \varsigma_{d,\eta} \varsigma_{\delta,\omega}   \frac{ \Ptx }{\sigma^2}.
\end{equation}
\end{corollary}
\begin{IEEEproof}
This result is proved in the same way as Corollary~\ref{cor:far-field-mMIMO} and Corollary~\ref{cor:far-field-relay}.
\end{IEEEproof}

The quadratic scaling with $N$ in \eqref{eq:SE-relaying-farfield} has been recognized in several recent works \cite{Wu2018a,Wu2019a,Basar2019a}, but without explaining that it only holds when the far-field approximation applies. Moreover, those papers noticed that SNR growth is faster than the linear scaling with $N$ observed for mMIMO receiver in \eqref{eq:SNR-mMIMO_planar-farfield} and for the mMIMO relay in \eqref{eq:SE-relaying-farfield}.
Although that implies that an IRS benefits more from increasing the array size, it does not mean that it will achieve a higher SNR when $N$ is large. Indeed, we already know from Proposition~\ref{prop:IRS-upperbound} and the subsequent discussion that this cannot happen in neither the far-field nor the near-field.

An instructive way of interpreting the $N^2$ scaling can be found by factorizing the far-field SNR in \eqref{eq:SNR:IRS2-far3} into two factors:
\begin{equation} \label{eq:SNR_IRS}
 \mathrm{SNR}_{\mathrm{IRS}}^{\mathrm{ff}}   = \hspace{-1.6cm} \!\!\!\!\!\!\overbrace{ \vphantom{\frac{  N\beta_{\vect{h}} \Ptx}{\sigma^2}} 
  N  \varsigma_{\delta,\omega}   }^{\leq 1, \textrm{ Fraction of reflected power reaching destination}}  \!\hspace{-2cm}\times \,\, \hspace{-0.2cm}\underbrace{  
N  \varsigma_{d,\eta}\frac{\Ptx}{\sigma^2 }   }_{=\,\mathrm{SNR}_{\mathrm{mMIMO}}^{\mathrm{ff}} }.
\end{equation}
The first factor contains one $N$-term and describes the fraction of power received at the IRS that also reaches the destination. Since this term is fundamentally upper bounded by one, this $N$-term describes a drawback rather than a benefit of using the IRS-type of relay (it is the fraction of power that is \emph{not} lost). The second factor in \eqref{eq:SNR_IRS} equals the far-field mMIMO SNR in \eqref{eq:SNR-mMIMO_planar-farfield} and its $N$-term represents the array power gain that is achieved when having a large array.

\begin{figure}[t!]
	\centering 
\begin{overpic}[width=1.07\columnwidth,tics=10]{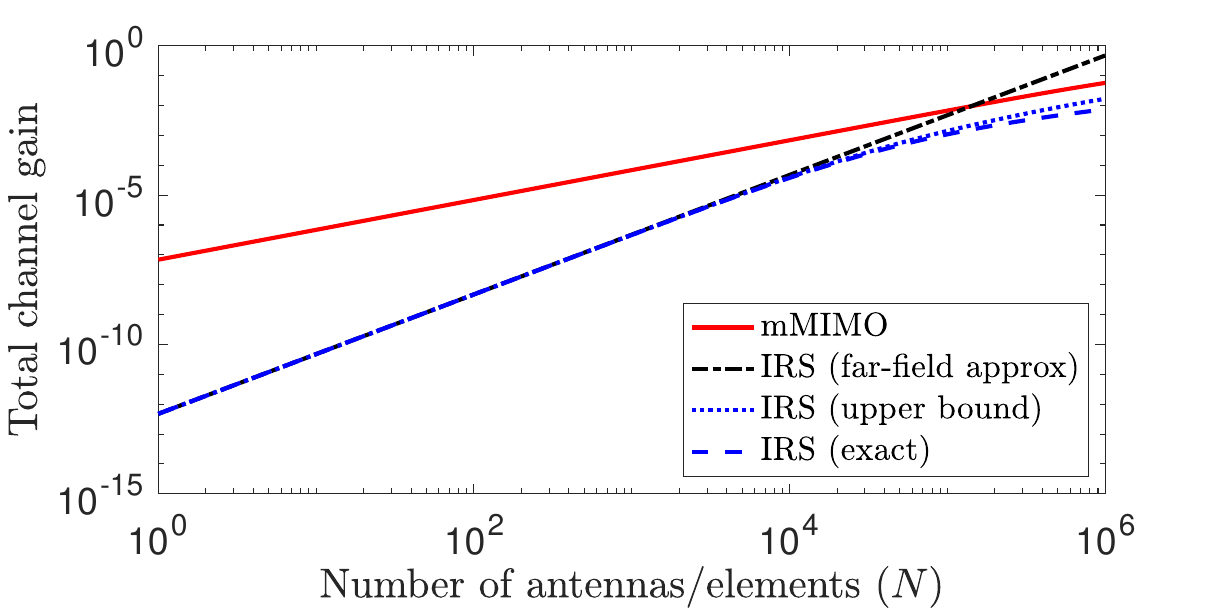}
	\put(38,43){\small Approximation is inaccurate}
	\put(77.5,44){\vector(1, 0){5}}
\end{overpic}  \vspace{-2mm}
	\caption{The total channel gain obtained with the mMIMO receiver and with the IRS-aided setup, for different number of antennas/elements $N$.
	The setup in Fig.~\ref{figure_figure_from_above} is considered with $d=25$\,m, $\eta=\pi/6$, $\delta=2.5$\,m, $\omega=-\pi/6$, $A=(\lambda/4)^2$, and $\lambda=0.1$\,m.}
	\label{figure_simulationIRSscaling}  
\end{figure}

To demonstrate these properties, Fig.~\ref{figure_simulationIRSscaling} shows the total channel gains obtained by the mMIMO receiver and the IRS-aided setup for a varying number of antennas/elements $N$.
We consider the setup in Fig.~\ref{figure_figure_from_above} with $d=25$\,m, $\eta=\pi/6$, $\delta=2.5$\,m, and $\omega=-\pi/6$. Each element in the array has area $A=(\lambda/4)^2$ with $\lambda=0.1$\,m.
We stress that in the IRS-aided setup, the destination is in the vicinity of the IRS.
The figure shows that the total channel gain grows as $N^2$ with the IRS and as $N$ with the mMIMO receiver, which is consistent with the respective far-field approximations derived in Corollaries \ref{cor:far-field-mMIMO} and \ref{cor:far-field-IRS}. Nevertheless, mMIMO provides a much larger channel gain for most values of $N$, which is consistent with Proposition~\ref{prop:IRS-upperbound}. The advantage remains asymptotically. The reason is that each element of the IRS acts as an isotropic scatterer, thus the IRS is a full-duplex relay that forwards the signal to the destination without amplifying it  \cite{Bjornson2019e}.
Even if the destination is close to the IRS, the far-field approximation in \eqref{eq:SNR:IRS2-far3} is accurate until the IRS has roughly $10^4$ elements. The upper bound in \eqref{eq:SNR:IRS2-bound} follows the exact curve closely, even for $N>10^4$, which is why we called it a tight bound.

\begin{remark} \label{remark:separability}
The upper bound in Proposition~\ref{prop:IRS-upperbound} contains the product of the total channel gain $\xi_{d,\eta,N}$  between the source and IRS and the total channel gain $\xi_{\delta,\omega,N}$ between the IRS and the destination.
This is the same structure as for the far-field SNR in \eqref{eq:SNR:IRS2-far3}, which has been analyzed in a series of previous works (e.g., \cite{Ozdogan2019a,Tang2019a,Garcia2019a,Ellingson2019a}).
However, the near-field behavior has not be analytically studied with the same rigor. The IRS was approximated as a specular reflector (i.e., an ideal mirror) in \cite{Basar2019a,Tang2019a,Yildirim2019a,Renzo2020a} and the channel gain in the near-field can then be made proportional to $1/(d+\delta)^2$ \cite{Renzo2020a}.
 This expression is different from the upper bound in Proposition~\ref{prop:IRS-upperbound}, where there are no terms that depend on both $d$ and $\delta$. 
The following conclusion can be made: If one can operate an IRS to get a channel gain of the kind in 
\cite{Basar2019a,Tang2019a,Yildirim2019a,Renzo2020a}, the SNR is likely not maximized by doing so. We return to this matter in Section~\ref{sec:geometric-interpretation}.
\end{remark}

We conclude this section by using the upper bound in Proposition~\ref{prop:IRS-upperbound} to study the asymptotic behavior of an IRS, particularly in the near-field.

\begin{corollary}[Asymptotic analysis] \label{cor:asymptotics-IRS}
As $N \to \infty$ with constant transmit power $\Ptx$, the SNR in the IRS setup is asymptotically upper bounded since
\begin{equation} \label{eq:SNR-IRS-limit}
\mathrm{SNR}_{\mathrm{IRS}}^{\mathrm{upper}} \to \frac{1}{9}\frac{ \Ptx}{\sigma^2}.
\end{equation}
If the transmit power is reduced with $N$ as $\Ptx = P/N^{\rho}$ for some constant $P>0$ and exponent $\rho>0$, then as $N \to \infty$ it follows that 
\begin{equation}
\mathrm{SNR}_{\mathrm{IRS}} =  \left( \sum_{n=1}^{N} |{h_n}||{g_n}| \right)^2 \frac{  P}{N^{\rho}\sigma^2} \to 0.
\end{equation}
\end{corollary}
\begin{IEEEproof}
The upper bound follows from the fact that $\xi_{d,\eta,N}, \xi_{\delta,\omega,N} \to 1/3$ as $N \to \infty$, which was also utilized in Corollary~\ref{cor:limit-mMIMO}. Since the channel gain is upper bounded, the SNR goes to zero if $\Ptx$ goes asymptotically to zero.
\end{IEEEproof}

This corollary shows, once again, that the asymptotic SE limit of any conventional power scaling law is zero. Nevertheless, we can expect the SNR in the IRS-aided setup to grow as $N^2$ for most practical array sizes. In those cases, it is also possible to reduce the transmit power as $1/N^2$ and keep the SNR constant. In agreement with Proposition~\ref{prop:IRS-upperbound}, Corollary~\ref{cor:asymptotics-IRS} also shows that an IRS-aided setup can never reach the same SNR as the mMIMO receiver for any common value of $N$. The difference remains as $N \to \infty$ since the limits are different.

\section{How Large IRS is Needed to Achieve the Same SNR?}
\label{sec:how-many-elements}

When looking for suitable use cases for the IRS technology, one needs to ask the question: How \emph{large} must the IRS be to achieve the same performance as with an active mMIMO receiver or a regenerative half-duplex mMIMO relay? To answer this question, we now let $N_{\mathrm{mMIMO}}$, $N_{\mathrm{relay}}$, and $N_{\mathrm{IRS}}$, denote the number of elements of the mMIMO receiver, the mMIMO relay, and the IRS, respectively.
We can then determine how many elements are needed in the IRS to achieve the same or higher SE than with the competing technologies.

\begin{corollary} \label{cor:N-comparison}
When operating in the far-field, the IRS case provides higher SE than the mMIMO receiver if
\begin{equation}
N_{\mathrm{IRS}} \geq  \sqrt{ \frac{N_{\mathrm{mMIMO}}  }{ \varsigma_{\delta,\omega}}}.
\end{equation}
Similarly, the IRS case provides higher SE than the half-duplex mMIMO relay if
\begin{equation}
N_{\mathrm{IRS}} \geq \sqrt{ \frac{\sigma^2}{ \Ptx \varsigma_{d,\eta} \varsigma_{\delta,\omega}} 
\sqrt{ 1 +
 N_{\mathrm{relay}} \frac{ \min \left( \Ptx  \varsigma_{d,\eta} , \Prelay  \varsigma_{\delta,\omega}  \right)}{\sigma^2}  } }.
\end{equation}
\end{corollary}
\begin{IEEEproof}
This follows from comparing the expressions in Corollaries \ref{cor:far-field-mMIMO}, \ref{cor:far-field-relay}, and \ref{cor:far-field-IRS}.
\end{IEEEproof}

\begin{figure}[t!]
	\centering 
	\begin{overpic}[width=1.07\columnwidth,tics=10]{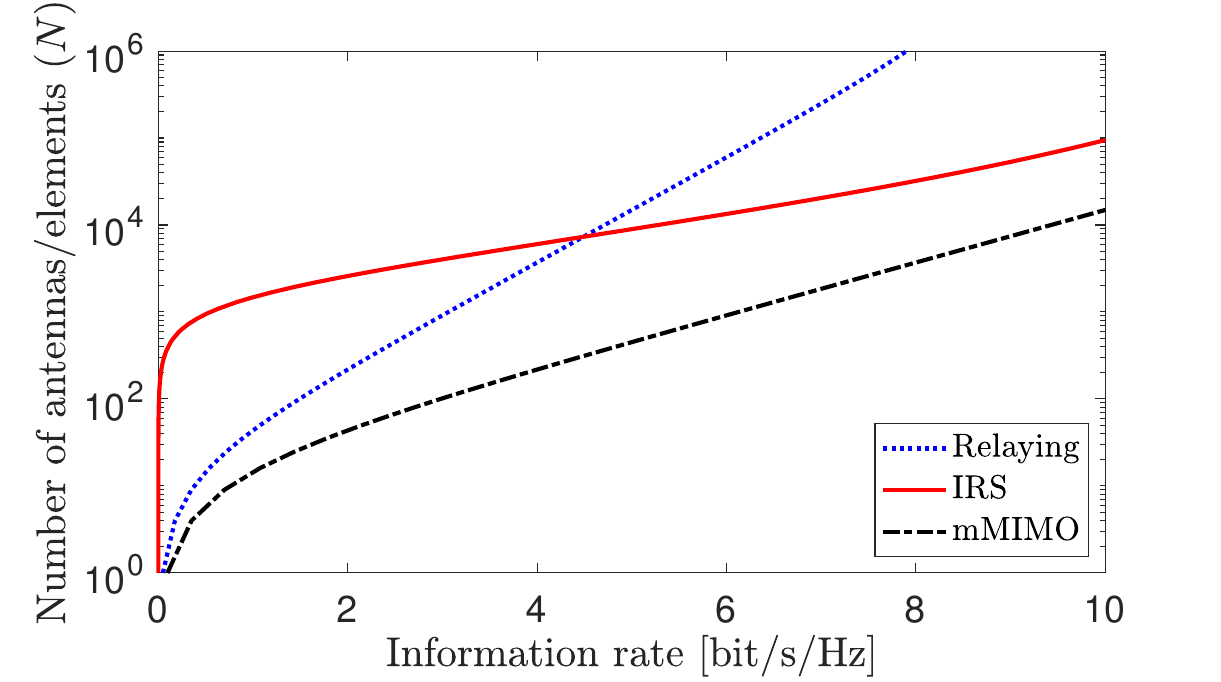}
\end{overpic} 
	\caption{The number of elements/antennas needed to achieve a given information  rate in the different setups. The same simulation parameters as in Fig.~\ref{figure_simulationIRSscaling} are considered with $\Ptx=\Prelay$.}
	\label{figure_simulationNcomparison}  
\end{figure}

By inserting values into the expressions in Corollary~\ref{cor:N-comparison}, Fig.~\ref{figure_simulationNcomparison} shows how many antennas are needed to achieve a particular SE in each of the three setups.
The same simulation parameters as in Fig.~\ref{figure_simulationIRSscaling} are considered with $\Ptx=\Prelay$. The first observation is that the IRS needs more than 100 elements before it provides an SE that is clearly above zero. After that, the number of elements grows more gracefully with the SE than for the half-duplex relay and mMIMO setups, since the SNR grows as $N^2$ for the IRS. However, it is only for SEs greater than 4.4\,bit/s/Hz that $N_{\mathrm{IRS}}<N_{\mathrm{relay}}$ in this example.
The IRS must always be larger than the mMIMO array to deliver the same SE. For example, $N_{\mathrm{mMIMO}}=100$ delivers 3\,bit/s/Hz, while $N_{\mathrm{IRS}}\approx 4000$ is needed to achieve the same SE. Since this example considers a 3 GHz carrier frequency, this corresponds to a mMIMO receiver that is $0.25 \times 0.25$\,m and an IRS that is $1.6 \times 1.6$\,m, thus it is practically possible deploy such an IRS even in an indoor environment. The difference in physical size reduces asymptotically but will not vanish, as proved in the previous section. If one would consider a different carrier frequency, the physical array dimensions remain the same but the number of elements required to build the array change.

\begin{remark}
Even if the IRS must be physically larger than the mMIMO counterparts to achieve a given SNR, it might still be practically preferable since the surface can be thin, integrated into existing construction elements, and, hopefully, cheap and energy-efficient. While the first generation of mMIMO technology is commercially available, the IRS technology is in its infancy which makes it impossible to quantify its cost and energy consumption \cite{Bjornson2020a}. It has been possible to build metasurfaces for many years but the IRS operation also requires real-time channel estimation and reconfigurability. This is an active research area \cite{Wu2019a,Taha2019a,Zheng2020} that has not converged to a mature solution yet and it is potentially the implementation of these functionalities that will dominate the hardware cost and energy consumption \cite{Bjornson2020a}.
\end{remark}

\section{Geometric Interpretation of Optimized IRS}
\label{sec:geometric-interpretation}

Some recent works model an IRS as a specular reflector or an ``anomalous mirror'' (i.e., a mirror with an unusual reflection angle) \cite{Basar2019a,Tang2019a,Yildirim2019a,Renzo2020a}
This basically means that the IRS reflects the incoming signal towards the destination as a flat and perfectly rotated plane mirror would do.
Under these conditions, the total channel gain of the IRS setup would converge to
\begin{equation} \label{eq:pathloss-mirror}
\varsigma_{d,\delta}^{\mathrm{mirror}} =  \left( \frac{\lambda}{4\pi (d+\delta)} \right)^2
\end{equation}
as the array size grows large and the near-field is considered. This asymptotic formula can be motivated by geometrical physics if one considers an equivalent setup where the destination is behind the mirror and the total propagation distance is $d+\delta$. Interestingly, the limit in \eqref{eq:pathloss-mirror} differs from the asymptotic upper bound derived in Corollary~\ref{cor:asymptotics-IRS}. More precisely, the two distances $d$ and $\delta $ appear in \eqref{eq:pathloss-mirror} in a joint factor $(d+\delta)^2$ and not within separate multiplicative factors as in Proposition~\ref{prop:IRS-upperbound}.
This reveals that an IRS that is optimized to maximize the SNR does not operate as a plane mirror.
 If a plane wave is impinging on a plane mirror, its specular reflection is also a plane wave. In contrast, if a plane wave is impinging on an IRS, each element will scatter a piece of the wave with a particular phase-shift. By optimizing the phase-shifts so that the $N$ scattered waves add constructively at the destination, the IRS effectively operates as a concave mirror that focuses the incoming wave at the point of the destination. The phase-shift optimization finds the SNR-maximizing curvature of the concave mirror and the IRS synthesizes such a mirror without actually changing its physical shape.

\begin{figure}[t!]
	\centering 
	\begin{overpic}[width=1.07\columnwidth,tics=10]{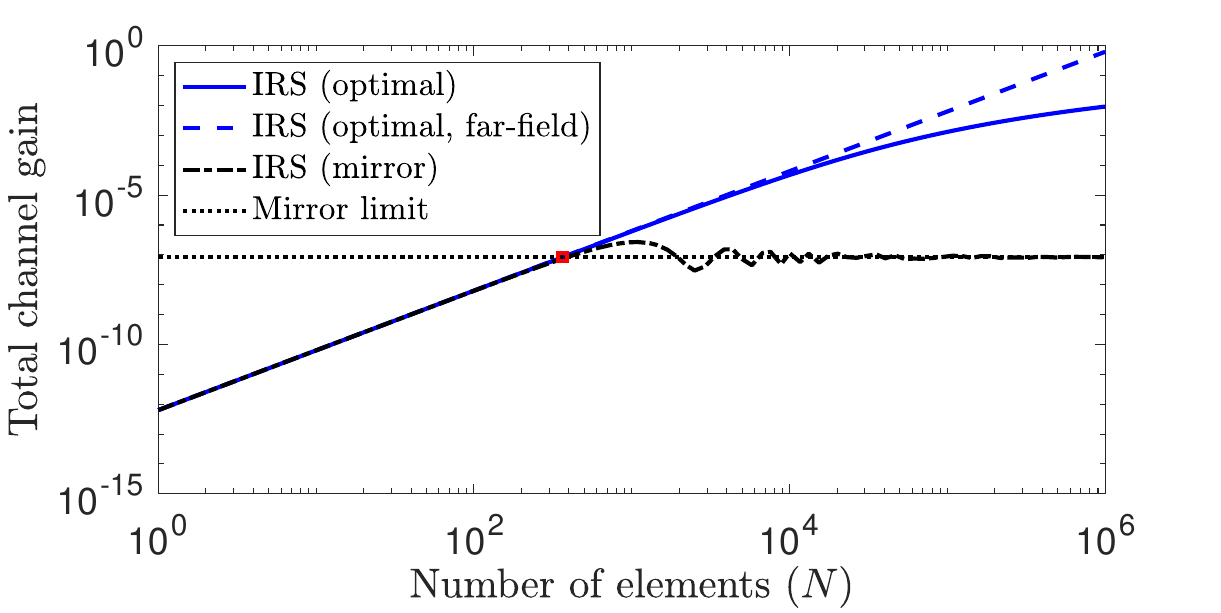}
		\put(44,19){\footnotesize The only point where an optimized}
		\put(44,16){\footnotesize   IRS operates as an anomalous mirror}
		\put(50,21.5){\vector(-0.5,1){3}}
\end{overpic} 
	\caption{The total channel gain obtained with the IRS-aided communication using either optimal phases or mirror-mimicking phases, for different number of elements $N$.
	The setup in  Fig.~\ref{fig:geometric-illustration}(b) is considered with $d=25$\,m, $\eta=0$, $\delta=2.5$\,m, $\omega=0$, $A=(\lambda/4)^2$, and $\lambda=0.1$\,m.}
	\label{figure_simulationIRSmirror}  
\end{figure}

Fig.~\ref{figure_simulationIRSmirror} demonstrates this in a setup where both the source and destination are centered in front of the array: $d=25$\,m, $\eta=0$, $\delta=2.5$\,m, $\omega=0$ following the notation in Fig.~\ref{fig:geometric-illustration}(b). In this case, a mirror-mimicking IRS has $\theta_n=0$ and $\mu_n=1$ for all $n$ and the corresponding total channel gain can be computed using \eqref{eq:SNR:IRS1} as $| \sum_{n=1}^{N}|{h_n}| |{g_n} | e^{-j(\phi_n+\psi_n)} |^2$. Fig.~\ref{figure_simulationIRSmirror}  also reports the total channel gain $( \sum_{n=1}^{N} |{h_n}| |{g_n} | )^2$ of an optimized IRS with $\theta_n = \phi_n+\psi_n$. Each element has area $A=(\lambda/4)^2$ with $\lambda=0.1$\,m.
There is no noticeable difference in Fig.~\ref{figure_simulationIRSmirror} for small IRSs because when the source and destination are in the far-field, focusing the incoming plane wave on a far-away point is approximately the same as mimicking a plane mirror that reflects the signal towards a point infinitely far away in the same angular direction. However, at around $N=360$, the channel gain of the mirror-mimicking IRS starts to converge to \eqref{eq:pathloss-mirror}, while the channel gain of the optimized IRS continues to increase. At $N=10^4$, the optimized IRS has a 500 times better channel gain than the mirror limit in \eqref{eq:pathloss-mirror}. 
We conclude that the SNR achieved by an \emph{optimized IRS} can generally not be described using the mirror limit; particularly not in the near-field since the dashed far-field approximation is accurate far beyond the point where optimized SNR surpasses the mirror limit. This conclusion is consistent with the results in \cite{Ellingson2019a}, which were derived by neglecting polarization effects. However, one can certainly use the mirror analogy to identify the approximately optimal phase-shifts when operating in the far-field \cite{Ozdogan2019a}.

By setting the far-field approximation in \eqref{eq:SNR:IRS2-far3}, for an optimized IRS, equal to the mirror limit in \eqref{eq:pathloss-mirror}, we obtain that
\begin{equation} \label{eq:exact-IRS-size}
NA = \left( \frac{1}{d} +\frac{1}{\delta} \right)^{-1} \lambda
\end{equation}
is the largest array area that a mirror-mimicking IRS can make use of in this example.\footnote{The expression is generalized in \cite[Eq.~(35)]{Ellingson2019a} to cases where the source and destination are located in different directions.}
This point is indicated by a square in Fig.~\ref{figure_simulationIRSmirror}. If the IRS is larger, the remaining area is essentially wasted on scattering signals in other directions. The same phenomenon appears when a person is looking into a large plane mirror and only sees his/her reflection in a small part of it.
 Hence, if one uses \eqref{eq:pathloss-mirror} as a proxy for the channel gain of an optimized IRS (e.g., as done in \cite{Basar2019a,Yildirim2019a}), then the results only hold when the IRS has exactly the area in \eqref{eq:exact-IRS-size} and the source/destination are centered in front of it.
If we change $\delta$ or $\lambda$, the curves in Fig.~\ref{figure_simulationIRSmirror} will be shifted in different directions, but the quantitative conclusions remain the same.
Note that a large area $NA$ is obtained in \eqref{eq:exact-IRS-size} as the wavelength is increase, thus a much larger area is needed to mimic a mirror in radio spectrum than in visible light.
In summary, the correct geometric interpretation of an optimized IRS is that it synthesizes the scattering off an optimally shaped concave mirror that can focus the incoming wave onto the point of destination.

\subsection{Reconfigurability Under Mobility}

The SNR-maximizing configuration focuses the reflected signal at the location of the destination, while the mirror-mimicking configuration forms a beam in the angular direction of the destination. 
One reason to consider the latter configuration is that only the angle must be known, thus the IRS must not be reconfigured if the destination moves along a trajectory where the angle is constant \cite{Renzo2020b}. However, even if the optimal focusing requires a continuous reconfiguration to be withheld under user mobility, one can also focus the signal at a point in the vicinity of the destination and keep this IRS configuration fixed as the destination moves.

\begin{figure}[t!]
	\centering 
	\begin{overpic}[width=1.07\columnwidth,tics=10]{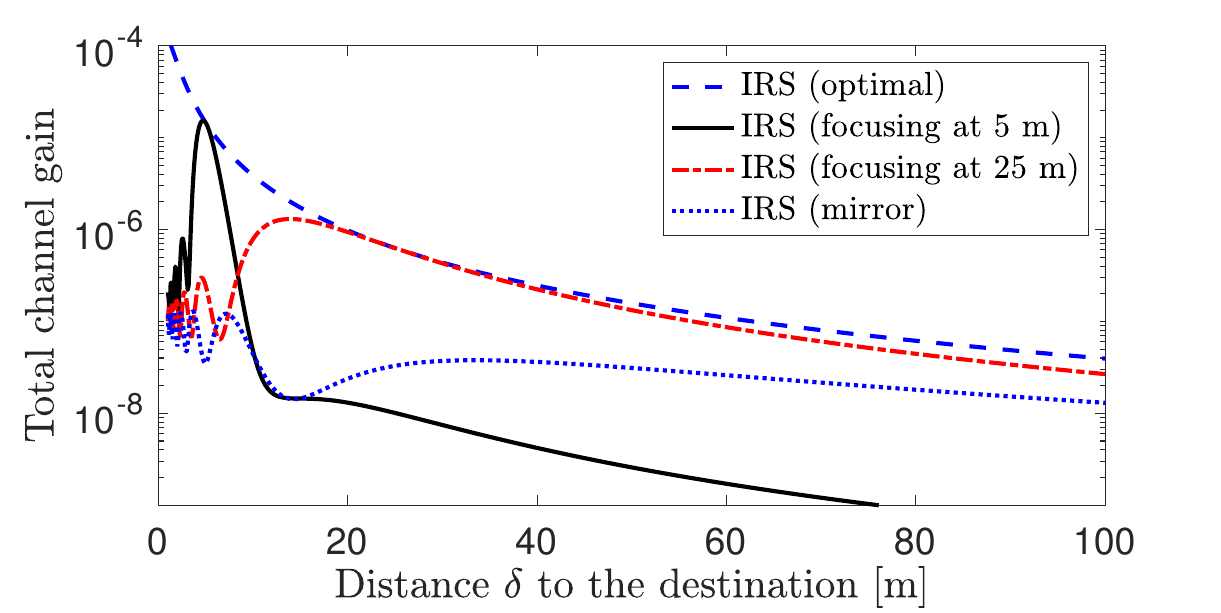}
\end{overpic} 
	\caption{The total channel gain when the destination is at different distances $\delta$ from the IRS. The SNR-maximizing configuration (utilizing the value of $\delta$) is compared with the mirror-mimicking case (only utilizing the angle $\omega$) and two cases where the signal is focused at a fixed point.}
	\label{figure_simulationIRSmirror3}  
\end{figure}

An example of this is provided in Fig.~\ref{figure_simulationIRSmirror3} for a setup where the source is located at $d=25$\,m from the IRS and the distance $\delta$ to the destination is varied. The figure shows how the total channel gain varies for $\delta \in [1,100]$\,m when using an IRS with $N=10^4$ elements. The setup in  Fig.~\ref{fig:geometric-illustration}(b) is considered with $d=25$\,m, $\eta=0$, $\omega=0$, $A=(\lambda/4)^2$, and $\lambda=0.1$\,m.
The upper curve represents the optimal SNR-maximizing configuration which requires reconfiguration of the IRS as the destination moves (i.e., a different $\vect{\Theta}$ for every value of $\delta$). There is a large gap to the dotted curve that represents the mirror-mimicking case that approximates specular reflection and beamforms the signal towards a point infinitely far away in the right angular direction. There are also two curves where the reflected signal is focused at a point that is either given by $\delta=5$\,m or  $\delta=25$\,m. These curves intersect with the optimal curve in the respective points but are otherwise below that curve, but the practical benefit is that the IRS is not reconfigured as the destination moves.
We notice that focusing at a nearby point (5\,m) leads to an array gain that is only obtained when the destination is close to the IRS, while the mirror-mimicking configuration outperforms it at larger distances. When the focus point is at an intermediate distance (25\,m), the achievable channel gain is larger than or approximately the same as in the mirror-mimicking case at all the considered distances and, particularly, preferable when the destination is at distances above 10\,m. The reason is that a specular reflector focuses the reflected signal at a point infinitely far away, thus it is only at very large distances (compared to the size of the IRS) that the mirror-approximation coincides with the SNR-maximizing configuration. Whenever there is side-information regarding the distance interval of where the destination might be, it is better to focus the signal at some point in that interval and then keep the configuration fixed even under mobility.

\section{Extension to General Propagation Setups}
\label{sec:extensions}

The analytical results have been derived in the free-space line-of-sight scenario, where the near-field behavior and asymptotic limits can be rigorously derived. Since the near-field behavior occurs when the propagation distance is comparable to the width/height of the array, it will mostly appear over short distances where the line-of-sight path exists. However, the channel can also contain additional scattered paths and the line-of-sight path can be partially blocked. Each such path can be modeled in the way described in Section~\ref{planar_arrays}, making the channel vector a summation of deterministic multi-path components arriving from different angles with different amplitude and phase.
Any such setup can be described by a pair of deterministic channel vectors $\vect{h}$ and $\vect{g}$ that satisfy the law of conservation of energy: $\|\vect{h}\|^2 \leq 1$ and $\|\vect{g}\|^2 \leq 1$.
For any such setup, it follows from \eqref{eq:SE-mMIMO} that the SE of the uplink mMIMO is
\begin{equation}
\mathrm{SE}_{\mathrm{mMIMO}} = \log_2 \left( 1+  \|\vect{h} \|^2 \frac{ \Ptx}{\sigma^2} \right).
\end{equation}
From \eqref{eq:SE-relaying}, the SE of the half-duplex mMIMO relay is
\begin{equation} \label{eq:SE-relaying-general}
\mathrm{SE}_{\mathrm{relay}} = \frac{1}{2} \log_2 \left( 1 + \min\left(\|\vect{h} \|^2 \frac{ \Ptx}{\sigma^2} , \|\vect{g} \|^2 \frac{ \Prelay}{ \sigma^2}  \right) \right).
\end{equation}
Furthermore, the SE of the IRS-aided setup is 
\begin{align} \notag
\mathrm{SE}_{\mathrm{IRS}} &= \log_2 \left( 1 + \frac{ \Ptx}{\sigma^2}
\left( \sum_{n=1}^{N} |{h_n}||{g_n}| \right)^2 \right) \\ & \leq  \log_2 \left( 1 + \| \vect{h} \|^2 \| \vect{g} \|^2 \frac{ \Ptx}{\sigma^2} \right) \label{SE_general_IRS}
\end{align}
where the upper bound follows from Proposition~\ref{prop:IRS-upperbound}. The SE will be strictly lower than that in \eqref{SE_general_IRS} if the IRS has a limited ability to control the phase-shift variables, but it seems that only a few discrete levels are sufficient to keep the SNR loss below 1\,dB \cite{Wu2020b}. In any case, \eqref{SE_general_IRS} remains a rigorous upper bound and the asymptotic limits are theoretically achievable.

If the arrays are equal-sized in all three setups, then the uplink mMIMO setup achieves the highest SE since $\| \vect{h} \|^2 \| \vect{g} \|^2  \leq \|\vect{h} \|^2$ for any practical channel setup. 
Whether the half-duplex mMIMO relay or the IRS-based relay achieves the highest SE depends on the channel model and transmit powers. Based on the previous results, we can expect the half-duplex relay to perform better when the arrays are small, while the IRS is preferable when the arrays are sufficiently large. The way to understand this is that the smaller pre-log factor of the half-duplex relay is particularly detrimental when the SNR is high, because then its higher SNR cannot compensate for it. If the IRS is physically larger than the mMIMO array, it can achieve the same or higher SNRs. No general relationship can be obtained.
When it comes to the asymptotic SE limits, the mMIMO and IRS-aided setups will give convergence to the same limit if $\| \vect{g} \|^2 \to 1$ as $N \to \infty$. However, this would require that the receiver captures all of the power that is scattered by the IRS, which might not occur in practice.

\section{Conclusions}
\label{sec:conclusion}

The limit of a large number of antennas has been studied in the multi-antenna literature for decades and is a core motivation behind the mMIMO technology. In this paper, we have noticed that previous asymptotic analyses have used channel models that are only accurate in the far-field, while the asymptotic limit can only be approached when operating in the near-field. Hence, the asymptotic SE behaviors and asymptotic power scaling laws in the existing literature can potentially be misleading. 
To determine when the asymptotic behaviors break down, we have derived a physically accurate channel gain expression for planar arrays, taking both polarization and near-field conditions (such as varying effective antenna areas) into account. 
We have used this model to revisit the power scaling laws and asymptotic limits in three MIMO setups: conventional mMIMO, half-duplex mMIMO relays, and IRS-aided communications.

The main observations are as follows. The total channel gains in the two mMIMO setups grow as $N$ in the far-field, where $N$ is the number of antennas/elements, while it grows as $N^2$ in the IRS setup. Numerical results showed that these behaviors are accurate even when the arrays have many thousands of elements/antennas, thus the classical scaling results are accurate in most practical deployments. 
However, the growth rate eventually tapers off when entering the near-field, and the channel gain converges to $1/3$ as $N\to \infty$ in the mMIMO setups, and is upper bounded by $1/9$ in the IRS setup. 
The near-field behavior begins when the width/height of the array is comparable (or larger) than the distance to the transmitter/receiver.
A consequence is that any power scaling law that lets the transmit power go asymptotically to zero will also lead to zero asymptotic SE.

The IRS will provably always achieve a lower SNR than the two mMIMO setups for any common value of $N$, despite the faster growth rate observed in the far-field. The reason is that one of the $N$-terms in the SNR accounts for the fraction of power that is lost in the IRS's reflection, thus it represents a drawback rather than a benefit.
However, if the IRS has a larger array size than in the mMIMO setups, it can achieve a higher SNR. This qualitative conclusion is previously known, but we have substantiated it by deriving exact expressions for when the breaking point occurs and shown that it appears in practically relevant cases.
Needless to say, the cost per element is lower with an IRS than in mMIMO, thus future work needs to consider if the total cost of the IRS technology will also be smaller.

By using the analytical expressions, we have proved that the SNR of an \emph{optimized} IRS contains the product of the channel gains from the source to the IRS and from the IRS to the destination, in both the near-field and the far-field. 
Previous works have interpreted the IRS as being an anomalous plane mirror (specular reflector) that can control the angular direction of the ``reflected'' signal, but we stress that  an optimized IRS synthesizes the scattering off a concave mirror that can also focus the signal on a point in the near-field. The optimal concave mirror is approximately plane when it is physically small and/or when the point is far away. In these cases, the optimal SNR does not match the ``sum-of-distances'' expression in \eqref{eq:pathloss-mirror} for an infinitely large plane mirror. In the near-field,  the IRS can achieve SNRs that are order-of-magnitude above that plane mirror limit.

The asymptotic analysis of this paper relies on a deterministic channel model, which is valid only for the considered propagation scenarios with fixed locations of the transmitter, receiver and arrays.
Unlike deterministic models, stochastic approaches are independent of a particular propagation environment and allow to model random reflections and scattering, and the channel fading they give rise to. Although there is no apparent reason to question the asymptotic findings of this paper under stochastic propagation conditions, it must be clear that the classical stochastic channel models cannot be used for the analysis of near-field behaviors since they do not capture the essential near-field propagation properties, such as the three key properties listed in Section~\ref{planar_arrays}. The research on this subject is generally open and nontrivial. In fact, the classical stochastic channel models do not well-reflect the physical properties of large arrays with a massive number of antennas in a compact space, not even in the far-field. A recent attempt to address this deficiency is~\cite{Pizzo2020}, where a spatially-stationary model for the small-scale fading in the far-field of non-isotropic random scattering environments is developed on the basis of a Fourier plane-wave spectral representation.

\appendices

\section{Proof of Lemma~\ref{lemma1}}
\label{app:proof-lemma1}
The proof proceeds in two steps. In the first step, we compute the channel gain of the $n$th antenna element using electromagnetic arguments (extending the work by \cite{Dardari2019a}) and show numerically how it can be tightly upper bounded when its area $a\times a$ is small (compared to the wavelength $\lambda$). In the second step, the upper bound on the channel gain is computed in closed form.

\subsection{Channel Gain Computation and Upper Bound}
Consider an \emph{elementary transmitting surface} with area $A_t$ and centroid located in $\vect{p}_t = [x_t,y_t,d]$.
The electric field ${\bf E}(\vect{p}_t, \vect{r})\in \mathbb{C}^{3}$ generated in a point $\vect{r} = [r_x,r_y,0]$, located in the Fraunhofer radiation region of the source, takes the form~\cite[Eq.~(4)]{Dardari2019a}
\begin{align}
{\bf E}(\vect{p}_t, \vect{r})  = {\bf G} (\vect{r} - \vect{p}_t){\bf J}(\vect{p}_t)
\end{align}
where ${\bf J}(\vect{p}_t) = J_x(\vect{p}_t)\hat{{\bf u}}_x+ J_y(\vect{p}_t)\hat{{\bf u}}_y + J_z(\vect{p}_t)\hat{{\bf u}}_z$ is the \emph{radiation vector}, which is measured in [A$\cdot$m] and determined by the surface current density. Notice that $\hat{{\bf u}}_x,\hat{{\bf u}}_y,\hat{{\bf u}}_z$ represent the unit vectors
in the $x,y,z$ directions. Also, ${\bf G} (\vect{r} - \vect{p}_t)\in \mathbb{C}^{3\times 3}$ is the Green function which, in the Fraunhofer radiation region of the source, is well-approximated as \cite[Eq.~(3)]{Dardari2019a}
\begin{align}\label{eq:green_function}
{\bf G} (\vect{r}) = -j \eta_0\frac{e^{-j \frac{2\pi }{\lambda}{||\vect{r}||}}}{2\lambda ||\vect{r}||}\left({\bf I}_3 - \hat{{\bf r}}\hat{{\bf r}}^{\Htran}\right)
\end{align}
with $\hat{{\bf r}} =  \frac{\vect{r}}{|| \vect{r}||}$ and $\eta_0$ denoting the impedance of free-space. This approximation is tight when the transmitter is beyond the reactive near-field of the receive antenna.

\begin{figure}[t!]
	\centering 
	\begin{overpic}[width=1.07\columnwidth,tics=10]{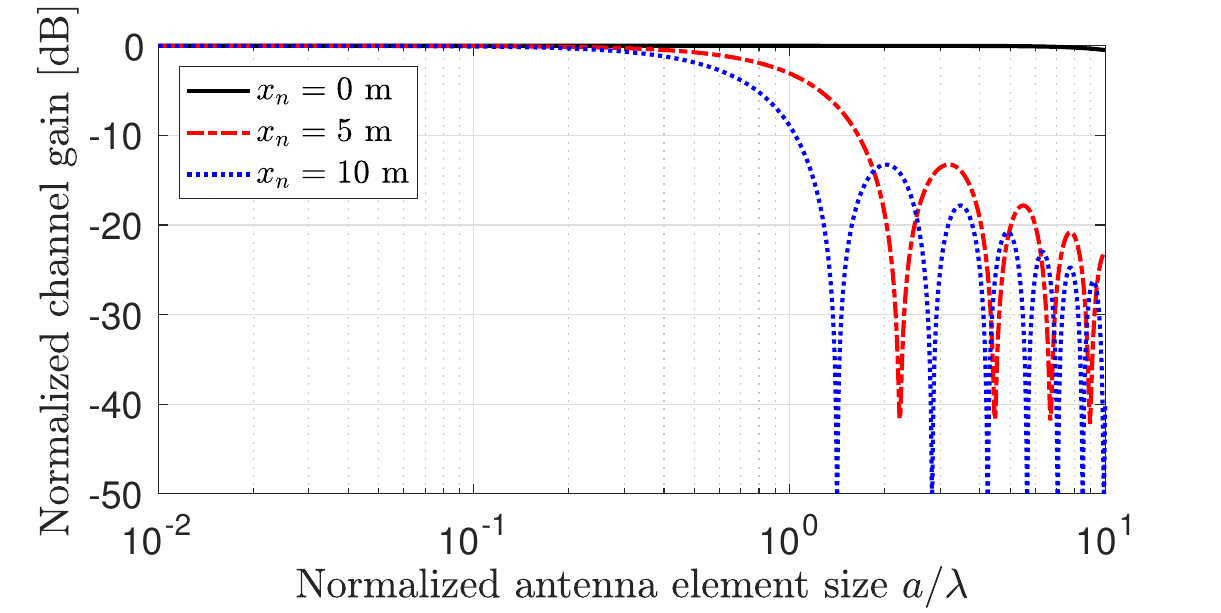}
\end{overpic} 
	\caption{Normalized channel gain $\left|h_n(\vect{p}_t)\right|^2/\zeta_{\vect{p}_t,\vect{p}_n}$ in dB as a function of $a/\lambda$ when ${\bf p}_n = (x_n, 0, 0)$ with $x_n=0,5$ and $10$\,m and ${\bf p}_t = (0, 0 , d)$ with $d=10$\,m.}
	\label{figure_normalized_channel_gain}  
\end{figure}

If only the $Y$ direction of ${\bf J}(\vect{p}_t)$ is excited at the source, then we have that ${\bf J}(\vect{p}_t) = J_y(\vect{p}_t)\hat{{\bf u}}_y$. The electric field reduces to 
\begin{align}
{\bf E}(\vect{p}_t, \vect{r})  & = {\bf G}_y (\vect{r} - \vect{p}_t)J_y(\vect{p}_t) 
\end{align}
where ${\bf G}_y (\vect{r}  - \vect{p}_t)={\bf G} (\vect{r} - \vect{p}_t)\hat{{\bf u}}_y$ is the second column of the Green function in \eqref{eq:green_function}. The complex-valued channel from the considered transmitting surface located in $\vect{p}_t$ to the receive point $\vect{r}$ located in the $XY$-plane is 
\begin{align}\label{eq:channel}
h(\vect{p}_t, \vect{r}) = |h(\vect{r} - \vect{p}_t)|e^{-j \frac{2\pi }{\lambda}{||\vect{r} - \vect{p}_t||}}
\end{align}
where \cite[Eqs.~(16) and (19)]{Dardari2019a}
\begin{align} \notag
|h(\vect{r} - \vect{p}_t)|^2 &=  A_t \overbrace{\frac{4}{\eta_0^2}||{\bf G}_y (\vect{r} - \vect{p}_t)||^2}^{\text{Power gain}}\overbrace{\frac{ ({\vect{r} - \vect{p}_t})^{\Ttran}\hat{{\bf u}}_z}{||{\vect{r} - \vect{p}_t}||}}^{\text{Projection on the $Z$ direction}} \\ &= \frac{1}{4 \pi}{\frac{ d \left( (r_x-x_t)^2 + d^2 \right)}{ \left( (r_x -x_t)^2 + (r_y-y_t)^2 + d^2 \right)^{5/2}}}\label{eq:channel_gain}
\end{align}
denotes the channel gain along the $Z$ direction (perpendicular to the array) where $\frac{ {\vect{r} - \vect{p}_t}}{||{\vect{r} - \vect{p}_t}||}$ denotes the pointing direction of the electric field and we have $A_t = \lambda^2/(4\pi)$ to represent the lossless isotropic antenna assumed in Lemma~\ref{lemma1}. Notice that~\eqref{eq:channel_gain} is measured in [m$^{-2}$]. If the $n$th antenna has the area $a  \times a$, the normalized (i.e., dimensionless) channel is
\begin{align}
h_n(\vect{p}_t) =  \frac{1}{a} \int_{x_n-a/2}^{x_n+a/2} \int_{y_n-a/2}^{y_n+a/2} h(\vect{p}_t, \vect{r}) \partial r_x \partial r_y.
\end{align}
The total channel gain is thus
\begin{align} \notag
\left|h_n(\vect{p}_t)\right|^2 &= \left|   \frac{1}{a} \int_{x_n-a/2}^{x_n+a/2} \int_{y_n-a/2}^{y_n+a/2} h(\vect{p}_t, \vect{r}) \partial r_x \partial r_y\right|^2 \\ &\leq \int_{x_n-a/2}^{x_n+a/2} \int_{y_n-a/2}^{y_n+a/2} \left|h(\vect{p}_t, \vect{r})\right|^2 \partial r_x \partial r_y   = \zeta_{\vect{p}_t,\vect{p}_n}\label{eq:channel_gain_approximated}
\end{align}
where the inequality follows from the Cauchy-Schwarz inequality and $\left|h(\vect{p}_t, \vect{r})\right|^2 = |h(\vect{r} - \vect{p}_t)|^2$ in \eqref{eq:channel_gain}. Fig.~\ref{figure_normalized_channel_gain} numerically evaluates the normalized channel gain $\left|h_n(\vect{p}_t)\right|^2/\zeta_{\vect{p}_t,\vect{p}_n}$ as a function of $a/\lambda$ when ${\bf p}_n = (x_n, 0, 0)$ with  $x_n =0,5$ and $10$\,m and ${\bf p}_t = (0, 0 , d)$ with $d =10$ m. The carrier frequency is $f=3$\,GHz. The results of Fig.~\ref{figure_normalized_channel_gain} show that $\zeta_{\vect{p}_t,\vect{p}_n}$ is a {tight upper bound} of $\left|h_n(\vect{p}_t)\right|^2$ for $a \le \lambda/4$ since the relative error in the channel gain is below $1$\, dB. An antenna element $a \le \lambda/10$ is needed to approach $0$\, dB. Motivated by this, we assume $a \le \lambda/4$ and replace $\left|h_n(\vect{p}_t)\right|^2$ with $\zeta_{\vect{p}_t,\vect{p}_n}$, which can be computed in closed form as shown next.

\subsection{Closed-form Expression of Channel Gain Bound} 
We will make use of the following primitive functions:
\begin{equation} \label{eq:identity1}
\int \frac{\partial x}{(x^2+a)^{3/2}}  = \frac{x}{a\sqrt{x^2+a}} + C
\end{equation}
\begin{equation} \label{eq:identity2}
\int \frac{\partial x}{(x^2+a)^{5/2}}  
= \frac{x}{3a(x^2+a)^{3/2}} + \frac{2x}{3a^2 \sqrt{x^2+a}} + C
\end{equation}
\begin{align} \label{eq:identity3} \nonumber
&\int \frac{\partial x}{(x^2+a) \sqrt{x^2+a+b}} \\ &= \frac{1}{\sqrt{ab}} \tan^{-1} \!\left( \frac{\sqrt{b}x}{\sqrt{a} \sqrt{x^2+a+b}} \right) + C
\end{align}
where $a,b$ are arbitrary scalars and $C$ is an arbitrary constant.

From~\eqref{eq:channel_gain}, it follows that $\zeta_{\vect{p}_t,\vect{p}_n}$ in \eqref{eq:channel_gain_approximated} requires to solve the following integral:
\begin{align} \notag
&\zeta_{\vect{p}_t,\vect{p}_n} \\ 
&= \frac{1}{4\pi} \int_{x_n-a/2}^{x_n+a/2} \int_{y_n-a/2}^{y_n+a/2} \frac{ d \left( (r_x-x_t)^2 + d^2 \right)   \partial r_x \partial r_y}{ \left( (r_x -x_t)^2 + (r_y-y_t)^2 + d^2 \right)^{5/2} } \label{eq:pathloss-integral} \\ \notag 
&= \int_{x_n-a/2}^{x_n+a/2} \int_{y_n-a/2}^{y_n+a/2}
\underbrace{ \frac{ d }{ \sqrt{ (r_x -x_t)^2 + (r_y-y_t)^2 + d^2 } }}_{\textrm{Reduction in effective area from directivity}} \\ &\quad \quad \quad \times \notag
\underbrace{\frac{  (r_x-x_t)^2 + d^2  }{ (r_x -x_t)^2 + (r_y-y_t)^2 + d^2  }}_{\textrm{Polarization loss factor}} \\ &\quad \quad \quad \times \notag \underbrace{\frac{  \partial r_x \partial r_y }{ 4\pi ( (r_x -x_t)^2 + (r_y-y_t)^2 + d^2 ) } }_{\textrm{Free-space pathloss}{}}
\end{align}
where $r_x,r_y$ are integration variables representing the location of the receive antenna. The contributions of the three fundamental properties when operating in the near-field of the array (i.e., the distance to the elements, the effective antenna areas, the loss from polarization) are clearly explicated. Next, we make the change of variables $\chi = r_x-x_t$ and $\upsilon = r_y-y_t$, so that \eqref{eq:pathloss-integral} becomes
\begin{align} \notag
&\frac{1}{4\pi}  \int_{x_n-a/2-x_t}^{x_n+a/2-x_t} \int_{y_n-a/2-y_t}^{y_n+a/2-y_t} \frac{ d \left( \chi^2 + d^2 \right)   \partial \upsilon \partial \chi}{ \left( \chi^2 + \upsilon^2 + d^2 \right)^{5/2} } \\ & =\frac{1}{4\pi} 
 \int_{x_n-a/2-x_t}^{x_n+a/2-x_t} \!\!
\left[
\frac{\upsilon d }{3(\chi^2 + \upsilon^2 + d^2 )^{3/2}} 
\right]_{y_n-a/2-y_t}^{y_n+a/2-y_t} \partial \chi \notag \\ &
+ \frac{1}{4\pi} 
 \int_{x_n-a/2-x_t}^{x_n+a/2-x_t} \!\! \left[ 
 \frac{2\upsilon d}{3(\chi^2 + d^2 ) \sqrt{\chi^2 + \upsilon^2 + d^2 }}
\right]_{y_n-a/2-y_t}^{y_n+a/2-y_t} \!\!\!\!\! \partial \chi \label{eq:pathloss-integral2a} \\   \notag
& =
\frac{1}{4\pi} \bigg(\sum_{y \in \mathcal{Y}_{t,n}  }
\!\! \int_{x_n-a/2-x_t}^{x_n+a/2-x_t}\!\!\!
 \frac{y d }{3(\chi^2 +  y^2+ d^2 )^{3/2}} \partial \chi \\  &
 +  \sum_{y \in \mathcal{Y}_{t,n}  }
 \!\!\int_{x_n-a/2-x_t}^{x_n+a/2-x_t}  \!\!\!\frac{2y d}{3(\chi^2 + d^2 ) \sqrt{\chi^2 + y^2+ d^2 }} \partial \chi\bigg)
\label{eq:pathloss-integral2b} 
\end{align}
by utilizing \eqref{eq:identity2}. The first integral in \eqref{eq:pathloss-integral2b} can now be computed using \eqref{eq:identity1}
 as
 \begin{align} \notag  &
  \int_{x_n-a/2-x_t}^{x_n+a/2-x_t}
 \frac{y d }{3(\chi^2 + y^2+ d^2 )^{3/2}} \partial \chi 
 \\ &
 = \sum_{x \in \mathcal{X}_{t,n} }  \frac{x y d}{(y^2+d^2) \sqrt{x^2+y^2 + d^2 }}. \label{eq:pathloss-integral2a-sol}
 \end{align}
 Moreover, the second integral in \eqref{eq:pathloss-integral2b} can be computed using \eqref{eq:identity3}
 as
\begin{align} \notag &
 \int_{x_n-a/2-x_t}^{x_n+a/2-x_t}  \frac{2y d }{3(\chi^2 + d^2 ) \sqrt{\chi^2 + y^2+ d^2 }} \partial \chi \\ &
=  \sum_{x \in \mathcal{X}_{t,n} } \frac{2}{3} \tan^{-1} \!\left( \frac{xy}{d \sqrt{x^2+y^2+d^2}} \right). \label{eq:pathloss-integral2b-sol}
\end{align} 
Substituting \eqref{eq:pathloss-integral2a-sol} into \eqref{eq:pathloss-integral2a} and  \eqref{eq:pathloss-integral2b-sol} into \eqref{eq:pathloss-integral2b} yield the final result in \eqref{eq:channel-gain-general-case}, after dividing the numerators and denominators by $d$.

\section{Proof of Corollary~\ref{cor:far-field-mMIMO}}
\label{app:proof-cor:far-field-mMIMO}

When $d \cos(\eta) \gg \sqrt{N A}$, it follows that $B+1 \approx 1$ and $2B+1 \approx 1$. We can then utilize that $\tan^{-1}(x) \approx x$ for $x \approx 0$  to approximate \eqref{eq:xi-mMIMO} as
\begin{align} \xi_{d,\eta,N} \approx \sum_{i=1}^{2} \frac{ B +(-1)^i \sqrt{B} \tan(\eta)  }{2 \pi \sqrt{ \tan^2(\eta) + 1 + 2(-1)^i \sqrt{B} \tan(\eta)} }.  \label{eq:xi-mMIMO-approx1}
\end{align}
Furthermore, we can utilize that $\sqrt{1+x} \approx 1 + x/2$ for $x\approx 0$ to approximate the denominator of \eqref{eq:xi-mMIMO-approx1} and obtain
\begin{align} \notag
\xi_{d,\eta,N} & \approx \sum_{i=1}^{2} \frac{ B +(-1)^i \sqrt{B} \tan(\eta)  }{2 \pi \sqrt{1+\tan^2(\eta)} \left( 1 + \frac{(-1)^i \sqrt{B} \tan(\eta)}{1+\tan^2(\eta)} \right)} \\ \notag
&
= \frac{ 2B - \frac{2B \tan^2(\eta) }{1+\tan^2(\eta)} }{2 \pi \sqrt{1+\tan^2(\eta)} \left( 1 + \frac{ \sqrt{B} \tan(\eta)}{1+\tan^2(\eta)} \right) \left( 1 - \frac{ \sqrt{B} \tan(\eta)}{1+\tan^2(\eta)} \right)} \\
& \approx \frac{B}{\pi (1+\tan^2(\eta))^{3/2}} = N  \underbrace{\beta_{d \cos(\eta)} \cos^3(\eta)}_{=\zeta_{d,\eta}}
\label{eq:xi-mMIMO-approx2}
\end{align}
where we first simplified the expression by writing the two fractions as a single fraction and then utilized that $1 - \frac{(-1)^i \sqrt{B} \tan(\eta)}{1+\tan^2(\eta)} \approx 1$ and finally that $1+\tan^2(\eta) = 1/\cos^2(\eta)$

\bibliographystyle{IEEEtran}

\bibliography{IEEEabrv,refs}

\end{document}